\documentclass[10pt,conference]{IEEEtran} 

\newif\ifARXIV
\ARXIVfalse
\ARXIVtrue

\usepackage{amssymb}
\usepackage{balance}
\usepackage{multirow}

\usepackage{changepage}

\newcommand{\MyInlineRequirement}[1]{\ensuremath{\mathbf{R}_{\mathbf{#1}}}}

\usepackage{graphicx}

\usepackage{tikz}
\usetikzlibrary{arrows.meta,positioning}

\usepackage[most]{tcolorbox}
\usepackage{xcolor}

\newtcolorbox{rawincidentbox}[2]{
  enhanced,
  breakable,
  colback=gray!18,
  colframe=gray!70,
  colbacktitle=gray!85,
  coltitle=white,
  boxrule=0.75pt,
  arc=1.2mm,
  left=6pt,
  right=6pt,
  top=2pt,
  bottom=2pt,
  before skip=5pt,
  after skip=5pt,
  fonttitle=\bfseries\small,
  title={#2 \hfill \normalfont\small \textit{#1}},
}

\newtcolorbox{defbox}[1]{
  enhanced,
  breakable,
  colback=violet!8,
  colframe=violet!60,
  boxrule=0.6pt,
  arc=1mm,
  left=6pt,
  right=6pt,
  top=3pt,
  bottom=3pt,
  before skip=6pt,
  after skip=6pt,
  fonttitle=\bfseries\small,
  title={Definition: #1},
}

\usepackage{enumitem}

\usepackage{amsmath}
\usepackage{booktabs}
\usepackage{todonotes}
\usepackage{comment}
\usepackage{url}
\usepackage{hyperref}
\usepackage{soul}

\usepackage[nameinlink]{cleveref}

\crefname{figure}{Fig.}{Figs.}
\Crefname{figure}{Fig.}{Figs.}

\crefname{table}{Table}{Tables}
\Crefname{table}{Table}{Tables}

\crefname{section}{}{}
\Crefname{section}{}{}
\creflabelformat{section}{#2\S#1#3}

\crefname{subsection}{}{}
\Crefname{subsection}{}{}
\creflabelformat{subsection}{#2\S#1#3}

\crefname{equation}{Equation}{Equations}
\Crefname{equation}{Equation}{Equations}

\usepackage{graphicx}

\makeatletter
\let\MYcaption\@makecaption
\makeatother

\usepackage[font=footnotesize]{subcaption}

\makeatletter
\let\@makecaption\MYcaption
\makeatother

\newcommand{\eg}{\textit{e.g.},\xspace}
\newcommand{\ie}{\textit{i.e.},\xspace}

\newif\ifDEBUG
\DEBUGtrue
\DEBUGfalse

\usepackage{xspace}

\newcommand{\StudyWeeks}{12\xspace}

\newcommand{\FieldNotesWords}{137,000\xspace}
\newcommand{\FieldNotesDescription}{over \FieldNotesWords words of contemporaneous field notes\xspace}
\newcommand{\NumIncidents}{88\xspace}

\newcommand{\RepoCodeLines}{2,140,456\xspace}

\newcommand{\RepoFiles}{11,659\xspace}

\newcommand{\ClaudeSubscriptions}{four\xspace}
\newcommand{\ClaudePlan}{Claude Max 20x\xspace}

\newcommand{\SalaryCostRounded}{50K\xspace}       
\newcommand{\InferenceCostRounded}{2K\xspace}     
\newcommand{\HostingCostRounded}{6K\xspace}       
\newcommand{\ClaudeCostRounded}{2K\xspace}        
\newcommand{\TotalDevCostRounded}{60K\xspace}     

\newcommand{\RepoCommits}{18,662\xspace}

\newcommand{\NumCustomLints}{577\xspace}   
\newcommand{\NumPredeployTests}{4{,}000\xspace}
\newcommand{\NumDesignDocs}{150\xspace}
\newcommand{\NumEpics}{30\xspace}

\newcommand{\ConcurrentAgents}{six to eight\xspace}
\newcommand{\CommitsPerDay}{500\xspace}

\newcommand{\ProductionKLOC}{420\,KLOC\xspace}              
\newcommand{\SupportApparatusMLOC}{1.16\,MLOC\xspace}        
\newcommand{\SupportRatio}{2.75$\times$\xspace}             
\newcommand{\SupportApparatusContents}{tests, lints, supporting documentation, and agent tooling\xspace}
\newcommand{\LoadBearingDocsKLOC}{247\,KLOC\xspace}  
\newcommand{\AgentInfraKLOC}{110\,KLOC\xspace}  
\newcommand{\ToolsKLOC}{162\,KLOC\xspace}  
\newcommand{\TestsKLOC}{405\,KLOC\xspace}  
\newcommand{\LintsKLOC}{238\,KLOC\xspace}  

\newcommand{\NumControlsCatalog}{41\xspace}                
\newcommand{\NumControlFamilies}{ten\xspace}               
\newcommand{\CommodityAnalyzers}{Roslyn (\texttt{AnalysisMode=All}), pyright (strict), ruff, and ESLint\xspace}
\newcommand{\NumCSharpTestMethods}{13{,}000\xspace}        
\newcommand{\NumPythonTestFiles}{1{,}500\xspace}           
\newcommand{\NumPropertyTests}{300\xspace}                 
\newcommand{\NumFuzzHarnesses}{90\xspace}                  

\begin{document}

\title{Cheap Code, Costly Judgment: A Theory of Agentic Software Engineering}
\title{Cheap Code, Costly Judgment: A Theory of Control Conversion in Agentic Software Engineering}
\title{Cheap Code, Costly Judgment: A Theory of Sustained Velocity in Agentic Software Engineering}
\title{Cheap Code, Costly Judgment: A Theory of Failure-Driven Control Hardening in Agentic Software Engineering}
\title{Cheap Code, Costly Judgment: A Failure Aware and Sustainable Agentic Software Engineering Theory}
\title{Cheap Code, Costly Judgment: A Theory of Control Hardening in Agentic Software Engineering}

\title{Cheap Code, Costly Judgment: Control Hardening as a Process for Agentic Software Engineering}
\title{Cheap Code, Costly Judgment: A Theory of Control Hardening in Agentic Software Engineering}
\title{Cheap Code, Costly Judgment: A Case Study on High-Quality, High-Velocity Software Engineering}
\title{Cheap Code, Costly Judgment: A Case Study on Governable High-Velocity Software Engineering}
\title{Cheap Code, Costly Judgment: How Agentic Velocity Becomes Governable Software Engineering}
\title{Cheap Code, Costly Judgment: A Case Study on Governance in Agentic Software Engineering}
\title{Cheap Code, Costly Judgment: A Case Study on Governance in Agentic Software Engineering}
\title{Cheap Code, Costly Judgment: A Case Study on Governable Agentic Software Engineering}

\ifARXIV
\author{
	\IEEEauthorblockN{James C. Davis}
	\IEEEauthorblockA{Purdue University\\
		davisjam@purdue.edu}
    \and
	\IEEEauthorblockN{Paschal C. Amusuo}
	\IEEEauthorblockA{Purdue University\\
		pamusuo@purdue.edu}
    \and
	\IEEEauthorblockN{Tanmay Singla}
	\IEEEauthorblockA{Purdue University\\
		tsingla@purdue.edu}\\
    \and
	\IEEEauthorblockN{Berk Çakar}
	\IEEEauthorblockA{Purdue University\\
		bcakar@purdue.edu}
	\and
	\IEEEauthorblockN{Kirsten A. Davis}
	\IEEEauthorblockA{Purdue University\\
		kad@purdue.edu}\\
}
\fi

\newcommand{\PA}[1]{%
  \ifDEBUG
    \todo[color=orange,inline]{PA: #1}%
  \fi
}

\newcommand{\BC}[1]{%
  \ifDEBUG
    \todo[color=cyan,inline]{BC: #1}%
  \fi
}

\newcommand{\JD}[1]{%
  \ifDEBUG
    \todo[color=green,inline]{JD: #1}%
  \fi
}

\newcommand{\TS}[1]{%
  \ifDEBUG
    \todo[color=blue,inline]{TS: #1}%
  \fi
}

\newcommand{\KD}[1]{%
  \ifDEBUG
    \todo[color=red,inline]{KD: #1}%
  \fi
}

\maketitle

\begin{abstract}
Generative AI is shifting software engineering from a practice organized around scarce implementation effort toward one organized around abundant, low-cost code production.
This shift changes the central engineering problem: not whether AI can generate useful code, but how engineers organize architectures, tools, evidence, and feedback loops so that AI-mediated development remains inspectable, correctable, and maintainable.

We study this problem through a first-person case study: a \StudyWeeks-week development effort in which a single expert software engineer used frontier AI coding agents to build a document accessibility remediation system.
The empirical record comprises \NumIncidents contemporaneous field notes,
\ProductionKLOC of production code, and \SupportApparatusMLOC of \SupportApparatusContents.
From this record, we develop a candidate middle-range theory of governance conversion, expressed as a process model explaining how high-velocity agentic implementation becomes governable.
The model explains how agentic implementation velocity surfaces recurring structural failure classes, and how engineering judgment sustains velocity by converting those failures into durable governance mechanisms.
In contrast to existing governance models that derive controls from known obligations, governance conversion explains how controls are discovered from failures that become visible only during agentic work.
We use our model to make testable predictions and to describe implications for software engineering research and practice.

\end{abstract}


\IEEEpeerreviewmaketitle


\section{Introduction}

Generative AI is changing the economics of software engineering~\cite{amodei2026policy}.
Recent coding agents let us trade human time for LLM tokens~\cite{jimenez2024swe}.
Software engineers must steer and validate unreliable, non-deterministic agents~\cite{he2025llm}.
The central problem is identifying engineering methods to make AI-mediated implementation governable at speed.

Prior work on AI-assisted software engineering has largely studied models, agents, and tools on bounded tasks such as
  code generation~\cite{dong2025survey},
  debugging~\cite{roy2024exploring},
  and
  repair~\cite{rondon2025evaluating}.
These studies show that agents can perform useful programming tasks and reshape developer workflows but provide limited evidence about end-to-end agentic software engineering processes.
Meanwhile, industry reports show that software engineers are using AI agents~\cite{anthropic2026compiler,cloudflare2026nextjs}, sometimes resulting in major production outages~\cite{moonka2026amazon_sev1_ai,ft2026amazon_ai_outages}.
Existing accounts do not explain how agentic development can sustain governable velocity:
 human-supervised workflows preserve oversight by making human attention the bottleneck,
 while multi-agent workflows accelerate implementation with underspecified quality control. 

We study this problem through a 12-week first-person case study~\cite{runeson2009guidelines,runeson2012case} of agentic development.
Here, the phenomenon of interest is not only the code produced by AI agents, but the situated sequence of judgments, controls, failures, and feedback loops through which agent-produced work was accepted, rejected, redirected, or converted into changes to the engineering environment. During the study period, a single expert software engineer used frontier AI coding agents to build a document accessibility remediation system. The empirical record includes 88 contemporaneous field notes, as well as design records, deployment data, and repository history. We analyze this record to learn: \textit{Can AI-mediated implementation be converted into governable software engineering progress?}

From this case, we develop a candidate middle-range theory~\cite{merton1968social} of governance conversion in agentic software engineering. The theory explains how high-velocity AI-mediated implementation can become governable software engineering progress. Its central process is \textit{failure $\rightarrow$ governance}: agentic velocity exposes structural failure classes, engineering judgment interprets those failures, and new governance mechanisms encode that judgment into the engineering environment to constrain subsequent agent work. 
The theory extends governance-centric accounts of agentic software engineering.
Prior governance work emphasizes \emph{ex-ante} governance: deriving controls from obligations known before agents act.
We identify a complementary \emph{ex-post} process: inducing controls from failures discovered during agentic work.
In this account, the scarce human work is not implementation-level review, but recognizing which failures reveal missing governance and converting them into architecture and controls.

This paper contributes:
\begin{itemize}
\item \textit{A process model of governance conversion in agentic software engineering,}
specifying its stages, outcomes, and influencing factors, along with testable propositions to inform future empirical research.

\item \textit{Empirical evidence from a sustained first-person case of agentic development:} a 12-week, richly instrumented case involving frontier AI coding agents, contemporaneous field notes, and repository history. 

\item \textit{A catalog of governance mechanisms for agentic software engineering,}
including recurring governance patterns and concrete examples from the development harness.
\end{itemize}

\textbf{Significance:}
This paper contributes an empirical and theoretical account of agentic software engineering from sustained firsthand participation in the development process.
The central lesson of the case is that AI-mediated implementation changes the bottleneck of software engineering: \textit{when implementation becomes abundant, sustainable progress depends on the governed engineering environment and on the human capacity to supply direction, interpretation, and abstraction}.
These findings inform several research agendas:
  empirical work to test and refine our theory;
  capability-building research on architectures, tools, and governance mechanisms for agent work;
  and educational work to train engineers to exercise architectural-contextual judgment at agentic speed.

\section{Background}
\label{sec:background}

This section motivates the need for a process account of governable agentic development.
\Cref{sec:background:genai}
defines agentic software engineering.
\Cref{sec:background:process-models} describes associated process models. 
\Cref{sec:background:gap} articulates the knowledge gap.

\subsection{Agentic Software Engineering}
\label{sec:background:genai}

Generative-AI coding systems have shifted from assistance-oriented tools toward agentic implementation systems.
Earlier systems primarily supported localized tasks, such as code completion~\cite{dong2025survey}, code optimization~\cite{peng2026sysllmatic}, code repair~\cite{rondon2025evaluating}, and unit test generation~\cite{schafer2024LLMUnitTest}. Although useful, these systems still required a human engineer to decompose the task, guide implementation, and integrate the suggested changes.
More recent systems are increasingly \emph{agentic}~\cite{sweagent,zhang-etal-2024-codeagent,autocoderover}: they can use repository context, invoke tools, and choose actions. This shift expands the scope of AI-supported development from localized edits to larger implementation tasks, including the construction of non-trivial software systems through natural-language interaction alone~\cite{deng2025swebenchproaiagents,ding2026nl2repobenchlonghorizonrepositorygeneration,yang2026programbenchlanguagemodelsrebuild}.


However, agentic software engineering also introduces new sources of failure and inefficiency. First, agent behavior is highly sensitive to its \emph{context}, including the task prompt, repository instructions (\eg \texttt{AGENTS.md}), tool descriptions, available files, and conversation history~\cite{mohsenimofidiContextEngineeringAI2026, liuLostMiddleHow2024}. Changes to this context can alter the assumptions available to the agent and therefore the implementation it produces~\cite{zhuoProSAAssessingUnderstanding2024}. 
Second, agents are stochastic systems operating over underspecified requirements. When constraints are omitted or ambiguous, the agent may infer requirements that were never intended, leading to unexpected or faulty behavior~\cite{yangWhatPromptsDont2026, yangAssessingImpactRequirement2026}. 
Third, agentic implementation introduces operational cost: each model interaction consumes tokens and latency~\cite{anthropic2026claudepricing,openai2026apipricing}. 
Thus, the shift to agentic software engineering creates a need for mechanisms that provide agents with better context and clearer constraints, so that increased implementation velocity does not come at the expense of quality, predictability, or cost.

\subsection{Process Models for Agentic SE}
\label{sec:background:process-models}

\begin{figure}[t]
\centering
\scriptsize
\setlength{\tabcolsep}{3pt}

\begin{tabular}{cc}
\textbf{(a) Velocity-Centric} & \textbf{(b) Oversight-Centric} \\[2pt]

\fbox{\begin{minipage}{0.45\linewidth}
\centering
\vspace{2pt}

\resizebox{0.92\linewidth}{!}{%
\begin{tikzpicture}[
  every node/.style={font=\footnotesize},
  role/.style={align=center, inner sep=2pt},
  arrow/.style={-{Latex[length=2mm,width=1.4mm]}, line width=0.45pt}
]
\node[role] (human) at (2.0,1.95) {\textbf{Human}};

\node[role] (deploy) at (0,1.0) {Deploy};
\node[role] (plan) at (2.0,1.0) {Planner};
\node[role] (dev) at (4.0,1.0) {Developer};

\node[role] (sec) at (0,0.25) {Security};
\node[role] (rev) at (2.0,0.25) {Reviewer};
\node[role] (test) at (4.0,0.25) {Tester};

\draw[arrow] (human) -- node[right, font=\footnotesize\itshape] {Goal} (plan);

\draw[arrow] (deploy) -- (plan);
\draw[arrow] (plan) -- (dev);
\draw[arrow] (dev) -- (test);
\draw[arrow] (test) -- (rev);
\draw[arrow] (rev) -- (sec);
\draw[arrow] (sec) -- (deploy);
\end{tikzpicture}%
}

\vspace{0pt}
\emph{Insufficient quality control}
\vspace{0pt}
\end{minipage}}
&
\fbox{\begin{minipage}{0.45\linewidth}
\centering
\vspace{2pt}

\resizebox{0.90\linewidth}{!}{%
\begin{tikzpicture}[
  every node/.style={font=\footnotesize},
  box/.style={align=center, inner sep=2pt},
  arrow/.style={-{Latex[length=2mm,width=1.4mm]}, line width=0.45pt}
]
\node[box] (human) at (0,0) {\textbf{Human}};
\node[box] (agent) at (3.0,0) {Agent completes\\bounded work};

\draw[arrow] (human.north east) to[bend left=28]
  node[above, font=\footnotesize\itshape] {prompt} (agent.north west);

\draw[arrow] (agent.south west) to[bend left=28]
  node[below, font=\footnotesize\itshape] {output} (human.south east);
\end{tikzpicture}%
}

\vspace{2pt}
\emph{Human attention gates progress}
\vspace{2pt}
\end{minipage}}
\\[30pt]

\textbf{(c) Governance-Centric} & \textbf{(d) Engineered Product} \\[2pt]

\fbox{\begin{minipage}{0.45\linewidth}
\centering
\vspace{2pt}

\resizebox{0.92\linewidth}{!}{%
\begin{tikzpicture}[
  every node/.style={font=\footnotesize},
  labelnode/.style={align=center, inner sep=2pt},
  envbox/.style={
  draw,
  rounded corners=3pt,
  align=center,
  inner sep=6pt
},
  agentbox/.style={draw, rounded corners=1pt, align=center, inner sep=3pt},
  arrow/.style={-{Latex[length=2mm,width=1.4mm]}, line width=0.45pt}
]
\path[use as bounding box] (-0.25,-0.45) rectangle (4.75,2.25);

\node[labelnode] (human) at (0.12,0.90) {\textbf{Human}};

\node[envbox, minimum width=3.45cm, minimum height=1.70cm] (env) at (3.30,0.90) {};
\node[labelnode] at (3.30,1.38) {Governed\\engineering environment};
\node[agentbox, minimum width=1.05cm, minimum height=0.42cm] (agent) at (3.30,0.40) {Agent};

\draw[arrow] (human.east) -- node[above, font=\footnotesize\itshape] {policies} (env.west);

\node[labelnode, font=\normalsize\itshape] at (2.25,-0.25)
  {Known policies become controls};
\end{tikzpicture}%
}

\vspace{2pt}
\end{minipage}}
&

\begin{minipage}{0.45\linewidth}
\centering
\vspace{2pt}

\resizebox{0.92\linewidth}{!}{%
\begin{tikzpicture}[
  every node/.style={font=\footnotesize},
  box/.style={draw, rounded corners=1pt, align=center, inner sep=4pt},
  source/.style={font=\footnotesize, align=center, inner sep=1pt},
  arrow/.style={-{Latex[length=2mm,width=1.4mm]}, line width=0.45pt}
]

\path[use as bounding box] (0.0,-0.25) rectangle (5.0,2.30);
\node[box, minimum width=2.55cm, minimum height=0.95cm] (prod) at (3.20,1.05)
  {\textbf{Product}\\software artifact};

\node[source] (fromA) at (0.55,2.18) {\textit{(a)}};
\node[source] (fromB) at (3.20,2.50) {\textit{(b)}};
\node[source] (fromC) at (0.50,1.05) {\textit{(c)}};

\draw[arrow] (fromA.east) -- (prod.north west);
\draw[arrow] (fromB.south) -- (prod.north);
\draw[arrow] (fromC.east) -- (prod.west);

\node[font=\normalsize\itshape, align=center] at (2.65,0.05)
  {Each model produces software artifacts.\\They differ in velocity/reliability tradeoffs.};
\end{tikzpicture}%
}

\vspace{2pt}
\end{minipage}

\end{tabular}

\caption{Dominant process models for agentic SE.
(a)~\textit{Velocity-centric} organizes agent labor to increase throughput, but leaves reliability mechanisms underspecified.
(b)~\textit{Oversight-centric} emphasizes quality, often gating velocity by putting human attention inside the implementation loop.
(c)~\textit{Governance-centric} defines a governed engineering environment before agent work begins, enforcing known (\textit{ex-ante}) obligations.
}
\label{fig:background:existing-theories}
\end{figure}
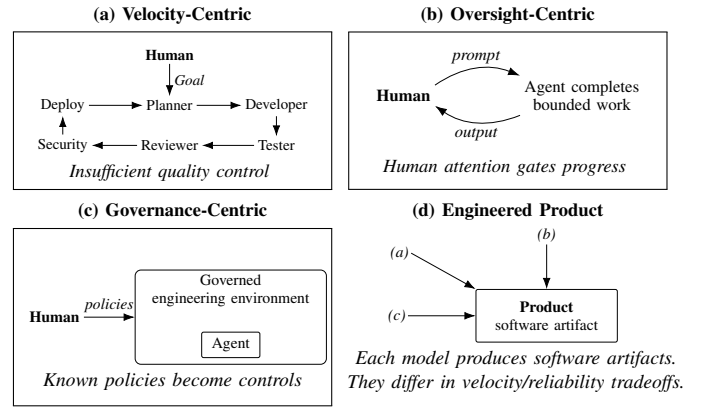

The main challenge of agentic software engineering is to ensure quality using fast, unreliable tools (\ie LLMs).
This has led to different accounts of how to organize agentic software engineering.
We organize prior accounts into three process models based on what they prioritize (\cref{fig:background:existing-theories}).

\subsubsection{Velocity-Centric~\cite{yegge2026gastown, qianChatDevCommunicativeAgents2024, hongMetaGPTMetaProgramming2024}}
This model gives agents autonomy, and may orchestrate multi-agent workflows to increase throughput.
For example, agents may be assigned engineering roles: planner, developer, and tester~\cite{yegge2026gastown}.
This increases velocity, but process resemblance is not control.
Agents can imitate the form of software work without producing quality output, and lack accountability, organizational context, and professional judgment.

\subsubsection{Oversight-Centric~\cite{bender2021parrots, chouBuildingSoftwareRolling2025, tangStudyDeveloperBehaviors2024}}
In this model, agents are useful but unreliable assistants, so human engineers remain near the implementation loop.
Humans prompt bounded work and inspect output.
This view recognizes that agentic output requires oversight.
Its limitation is that the control mechanism is human attention. 
At high implementation volume, oversight becomes the throughput bottleneck.
A related line is trace-based agent improvement, in which execution histories are manually or automatically analyzed to distill reusable lessons into skills and harness changes~\cite{shinn2023reflexion,barke2026agentrx,ni2026trace2skill,lin2026agenticHarnessEngineering,chen2026harnessfix}.

\subsubsection{Governance-Centric~\cite{rahimi2026agentic,kaptein2026runtime,koch2026governance}}
This emerging model holds that agents create product, organizational, and legal risks as they act~\cite{kraprayoon2025ai}.
The literature proposes policy-to-control methods.
Its techniques start from known \textit{ex~ante} obligations (\eg regulations) and deduce corresponding agent controls~\cite{ait2025towards}, \eg as runtime guardrails.

\begin{defbox}{Agentic Governance}
\small
We use \emph{agentic governance} in the operational sense of recent agent-governance work~\cite{kaptein2026runtime,koch2026governance,ait2025towards}: the mechanisms by which autonomous agents can be kept productive while holding the cost of their failures within acceptable bounds.
A governance mechanism may be a \textbf{control} (detecting or containing failures) or \textbf{architecture} (eliminating failures by construction).
\end{defbox}

\subsection{Knowledge Gap}
\label{sec:background:gap}

Together, these bodies of work expose a gap in software engineering process knowledge.
Conventional software engineering has processes for controlling change throughout the development lifecycle~\cite{royce1970managing,beck2001agilemanifesto}. 
Agentic software engineering changes the calculus: the potential rate of change skyrockets.

The open question is whether AI-mediated implementation can remain governable at speed. 
The dominant process models trade off velocity and quality.
Although the emerging governance approach is promising, we do not yet know what controls become necessary beyond ex-ante ones, nor how engineers revise the control environment when agent-produced failures reveal obligations not specified in advance. Existing work on control induction improves agent skills and harnesses from traces, but does not explain how observed failures harden the broader engineering environment.
We call this process \textit{governance conversion}: the conversion of observed agentic failure patterns into explicit, durable governance that constrains subsequent agent work. This paper studies governance conversion during a sustained agentic development effort.

\section{Methodology}
\label{sec:method}

\subsection{Research Question}
\label{sec:method:RQ}

Given this gap in process knowledge, we ask:

\vspace{0.02cm}
\begin{adjustwidth}{-0.5em}{0pt}
\begin{quote}
\textbf{RQ:} \textit{Through what process can high-velocity agentic
implementation be converted into governable progress?}
\end{quote}
\end{adjustwidth}


\subsection{Study Design}
\label{sec:method:StudyDesign}

We address this question through a case study. This design follows from the character of the research question. \textit{First}, case study research is appropriate when a phenomenon must be understood in its real-world context, especially when the phenomenon and context are difficult to separate~\cite{yin2018case,runeson2009guidelines}. Here, the object of study is not AI-generated code, but the process through which an engineer delegated work to agents, interpreted their outputs, and adapted the development environment over time. \textit{Second}, case studies are suitable for theory development: they can identify constructs, relationships, and candidate mechanisms before those constructs are mature enough for population-level measurement~\cite{george2005caseStudies,eisenhardt1989building,eisenhardt2007theory}.

We used a first-person case design because it made the target process unusually observable.
This design provides access to situated decisions and co-analyzable engineering artifacts that are difficult to recover through post-hoc reconstruction~\cite{seaman1999qualitative,dittrich2007cooperative,runeson2009guidelines}.
The bounded case is the 12-week development effort described in~\cref{sec:case}:
  one expert engineer (``\textit{Subject}'') using frontier coding agents to develop a greenfield software system.
This case exposed repeated decisions about what to delegate, what to trust, what to inspect, and how to revise the engineering environment as failures appeared.

Our study design supports theory building rather than prevalence estimation.
A single case cannot establish prevalence. 
Its value is analytic: it provides a richly instrumented account from which constructs, candidate mechanisms, and failure modes can be identified~\cite{yin2018case,george2005caseStudies}.
Once specified, other empirical methods can test such constructs.
We contribute the prerequisite step, by characterizing a sustained agentic development process to articulate those constructs.

\subsection{Data Sources and Analysis}

\subsubsection{Data sources}
We draw on two sources of evidence.
First, the Subject maintained contemporaneous field notes during the 12-week study period, written as 88 episode-bounded entries.
Second, we analyzed engineering artifacts from the project repository, comprising \RepoCommits commits and approximately 1.6~million lines of active artifacts. 
The two sources triangulate: the field notes record what the Subject perceived as salient, while the repository history is an independent trace of what actually changed---we used it to trace the governance mechanisms and themes that emerged from the field notes back to the commits that underpinned them.

\subsubsection{Incident construction}
We analyzed the 88 episodes using structured analytic memoing~\cite{saldana_coding_2013}.
Each episode was treated as a \textit{critical incident}~\cite{flanagan_critical_1954,butterfield_fifty_2005,gremler_critical_2004}, an event that the Subject judged salient enough to record during the study period, including failures, architectural decisions, and changes in engineering process.
For each incident, we made an analytic memo recording the triggering condition, relevant inputs, the Subject’s interpretation, the engineering response, the observed outcome, and the apparent effect on subsequent agent-mediated work.
\textit{Below, the beginning of a May 28 incident:}

\begin{rawincidentbox}{05-28-2026}{What kind of code \emph{can't} we write with Claude?}
\small
\textit{Meta-coding with Claude. This idea saved about a day of Sonnet agents working steadily---ask me how I know; hint: not the first such lint we developed with a big blast radius and very mechanical changes---and built naturally on the expertise the codebase had developed in making lints. The insight needed was to jump from ``Claude can code'' (done) to ``Claude can analyze code'' (done, a bit harder for Claude) to ``Claude can write analysis code to refactor automatically'' (a jump---Claude had much more trouble with this than the other two kinds). ...}
\end{rawincidentbox}

\newcommand{\CodebookNumIterations}{11\xspace}

\subsubsection{Codebook development}
The Subject led codebook development through iteration over the incident corpus.
In each pass, the Subject tested candidate taxonomies and revised category boundaries to better fit the data.
LLMs were used as qualitative-analysis instruments: normalizing field notes, comparing incidents against candidate codebooks, identifying provisional themes, and locating corroborating repository evidence.
LLM-generated summaries, labels, and mappings were treated as analytic aids, with interpretive authority remaining with the author team.
The codebook stabilized in \CodebookNumIterations iterations.

As a validity check, a second author independently re-coded a stratified sample of 10 incidents, together with the emergent themes, working from the codebook without the Subject's labels.
Agreement was high: 10/10 on incident class, 6/7 on category, and 5/7 on the third coding layer.
The second author judged all 10 incident summaries supported by the underlying evidence.
The few category and third-layer differences were deemed boundary cases rather than errors.

\subsubsection{Reflexivity}
Because the Subject was both practitioner and analyst, the account is shaped by his situated judgment.
Following first-person research practice~\cite{finlay2002reflexivity,tracy2010qualitativeQuality}, we embedded several reflexive moments of \emph{forced articulation}.
Across 6 weeks of the study period, the Subject prepared and delivered four talks about the project to institutional, research, and teaching audiences.
External response was uncritical, so the value of these checkpoints lay not in peer feedback but in the articulation itself.
Repeatedly explaining the work to outsiders forced the Subject to convert his intuitions into explicit constructs.
This articulation informed the codebook and final interpretation of the case.

\subsubsection{Nature of result}
Our analysis is interpretive and theory-building (\cref{sec:theory}).
We contribute a process model and accompanying theory that future work can refine, challenge, and test.

\subsection{Researcher and Participant Role}
\label{sec:Method:Participant}

This is a first-person case study of expert agentic software engineering.
The primary practitioner in the case is called the \textit{Subject}:
  a software engineering professor at a U.S. research university who is also an author of this paper.
The other authors' role was to help analyze the empirical record, challenge interpretations, and refine constructs. 

The Subject needed engineering expertise to undertake the case. 
The Subject has sixteen years of professional experience, as a software engineer (5 years), PhD student (5 years), and professor (6 years).
His engineering experience was prior to the GenAI era, involving his development of $\sim$100,000 lines of code in C/C++, Perl, and Python.
His research has contributed to major software infrastructure, \eg programming language runtimes.
His group has published several agentic systems. 

\section{Case Description}
\label{sec:case}

This section defines the case boundary and engineering context needed to interpret the incident analysis.

\subsection{Regulatory Context and System Requirements}
\label{sec:case:context-requirements}

The case was in the context of U.S. public-sector digital-accessibility regulation. 
This sector is governed by the Americans with Disabilities Act (ADA)~\cite{ada1990}.
In 2024, the U.S. Department of Justice updated the ADA guidance for digital content~\cite{doj_titleii_finalrule},
requiring that the digital content of public entities communicate ``\textit{equally effective[ly]}'' to people with and without disabilities~\cite{doj_effective_communication}, operationalized as WCAG~2.1~Level~AA~\cite{wcag21}.
The obligation applies to electronically distributed materials, \eg slide decks (PPTX) and readings (PDF).
For large entities (\eg~NC State, Penn State), DOJ set an April 2026 compliance deadline.
Many public universities, including the authors', did not license commercial remediation services.\footnote{The reason may be cost: communication with university acquisition staff indicates pricing of $\sim$\$3 per page or slide, or \$10-15K for a semester-long course's slides and readings; universities have thousands of courses.}
Instructors remediated by hand. 


The ADA informs the system requirements \MyInlineRequirement{i}:
\begin{enumerate}[leftmargin=*, itemsep=0pt, topsep=2pt]
\item \MyInlineRequirement{1}:
An automated system's \textit{inputs} are educational documents, \eg Office (PowerPoint, Word, Excel) and PDF. 
\item \MyInlineRequirement{2}:
Those inputs must be \textit{transformed for accessibility}. 
Some WCAG 
  transformations are syntactic and deterministic, \eg text size and color contrast (Criterion 1.4.3).
Other transformations are semantic.
Some can be done by dispatch to a multimodal model, \eg ``alt-text'' (\textit{Criterion 1.1.1: non-text content needs an equivalent text alternative}).
Others require document-structure remediation that can use generative AI as an algorithmic component, \eg
  that (\textit{Criterion 1.3.1}) information, structure, and relationships conveyed through presentation must be programmatically determinable or available in text,
  and
  (\textit{Criterion 1.3.2}) content whose sequence affects meaning must have a programmatically determinable reading sequence.
Accessibility itself is an underspecified construct~\cite{kelly2021accessibility}, dependent on audience, instructional purpose, and faculty preference.
\item \MyInlineRequirement{3}:
System outputs must be \textit{auditably traceable} to compliance requirements. The system must record what accessibility requirement motivated each change, and what evidence supports the resulting document state.
This excludes the black-box whole-document use of frontier models.
\item \MyInlineRequirement{4}:
The system should \textit{scale}: to thousands of faculty at the authors' institution, and preferably nationwide.
\end{enumerate}

This setting posed two engineering challenges.
First, the time pressure made a conventional (non-agentic) engineering approach infeasible.
Second, the Subject could not assemble a remediation pipeline from existing components.
The document-accessibility ecosystem contains many checkers to facilitate manual remediation~\cite{verapdf_home,adobe_acrobat_accessibility,libreoffice_accessibility,microsoft_accessibility_checker}.
There are no mature open-source tools for end-to-end \textit{remediation} of documents.
Hence, the Subject and his AI agents operated in a weakly templated domain.
The system required file-format manipulation, reasoning about accessibility semantics, expressing changes as sequences of auditable generative-AI queries, validation, and deployment in a space with few mature remediation exemplars.

Together, the deadline pressure, weak tool ecosystem, semantic remediation burden, auditability requirement, and SaaS deployment target made the project a substantial engineering effort ---
and a suitable setting for our research question.


\subsection{Engineering Approach}
\label{sec:case:agentic-environment}

\begin{figure}[t]
  \centering

  \begin{subfigure}[t]{0.49\columnwidth}
    \centering
    \includegraphics[width=\linewidth]{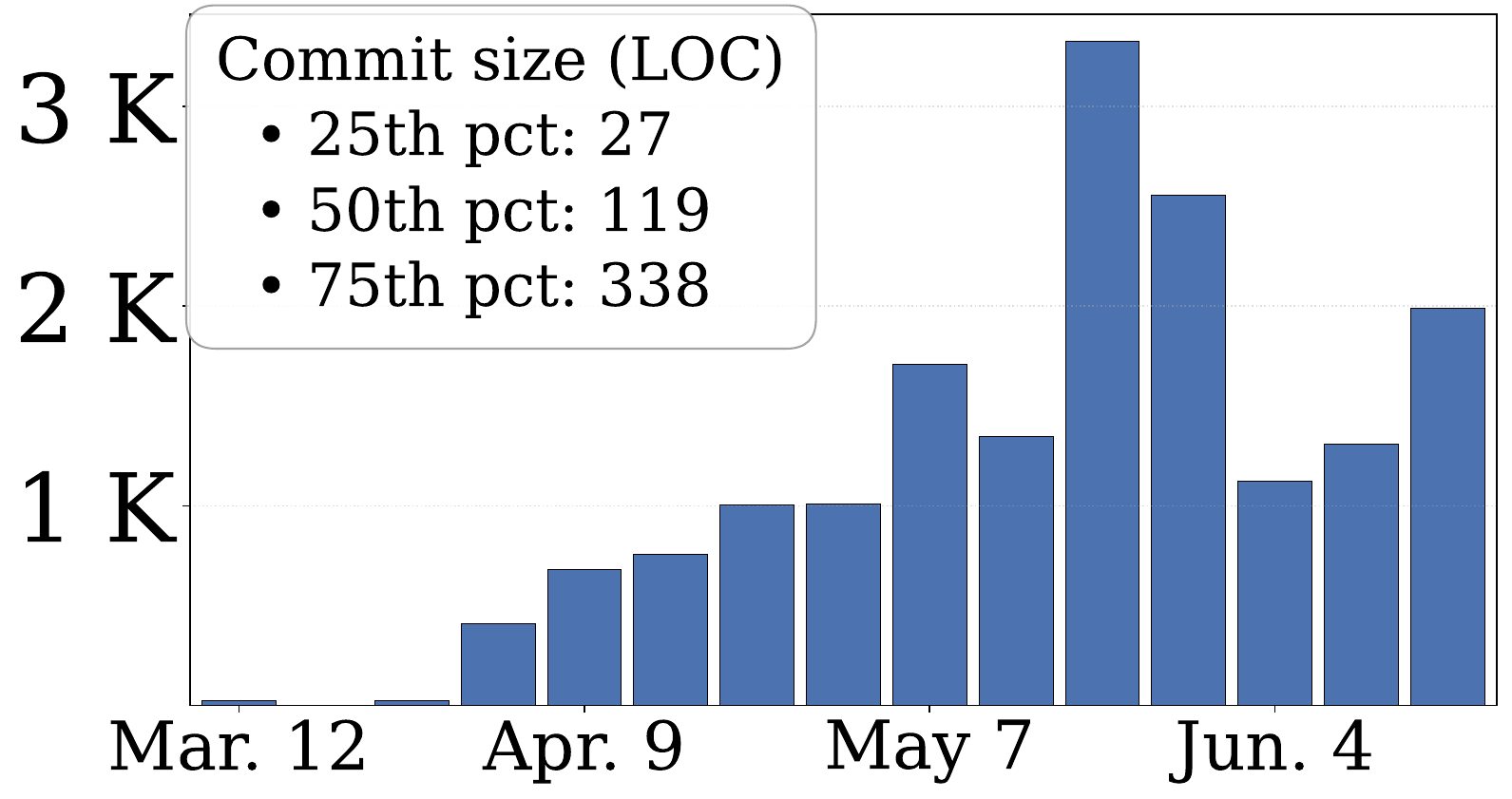}
    \caption{Commits per week.}
    \label{fig:repo-activity-commits}
  \end{subfigure}
  \hfill
  \begin{subfigure}[t]{0.49\columnwidth}
    \centering
    \includegraphics[width=\linewidth]{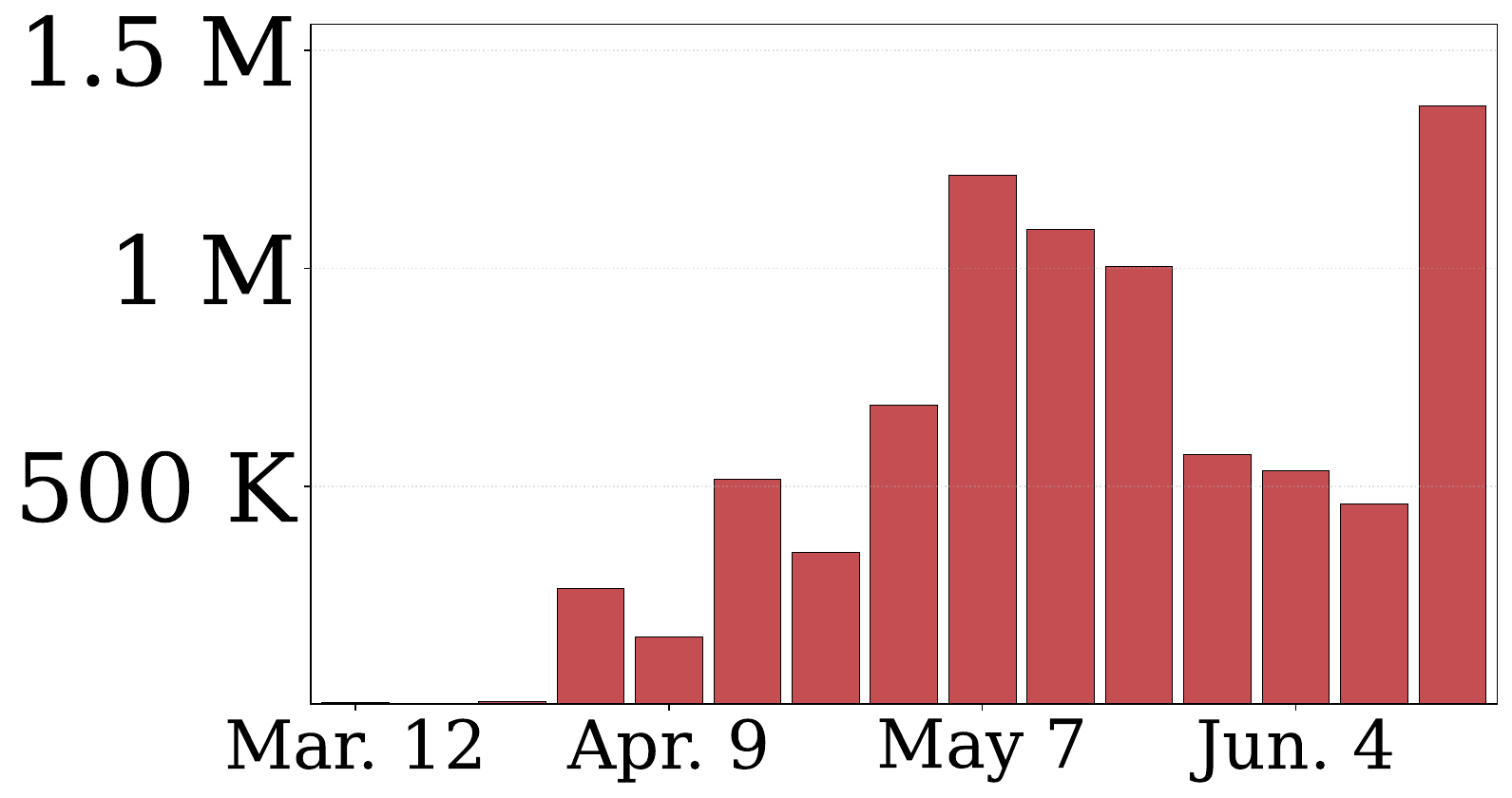}
    \caption{Lines changed per week.}
    \label{fig:repo-activity-loc}
  \end{subfigure}

  \caption{
  Repository activity over the case-study period.
  Velocity rose through the MVP stages and fell during hardening.
  }
  \label{fig:repo-activity}
\end{figure}

\begin{figure*}[t]
    \centering
    \includegraphics[width=\textwidth]{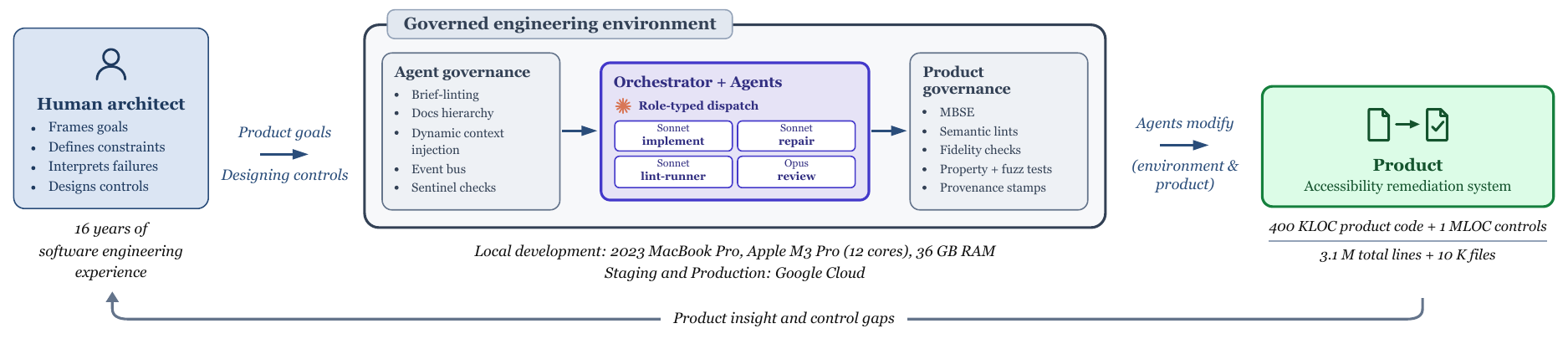}
    \caption{
    Development workflow used in the case study.
    The Subject supplied product goals and governance updates.
    Within the governed engineering environment,
   \textit{Agent governance} constrained agent behavior,
   and
   \textit{Product governance} constrained candidate changes before they entered the software artifact.
   Product insights and failures provided feedback for the Subject.
   The Subject converted that feedback into stronger governance to sustain development velocity.
   }
    \label{fig:case-process-overview}
\end{figure*}

The Subject undertook this project as his primary work activity for the 12-week study period.
\cref{fig:repo-activity} shows the achieved implementation velocity,
 and
\cref{fig:case-process-overview} shows his process.

\subsubsection{Iterative Capability Development}
\label{sec:case:agentic-environment:process}

Given the greenfield nature of the project, the lack of mature approaches, and the uncertainty regarding feasibility and user needs, the Subject selected an iterative development process, with six stages:

\begin{enumerate}[leftmargin=*, itemsep=0pt, topsep=2pt,label=\alph*)]
\item \textit{Stage 1: Feasibility Probe (Mar.~2026)}:
  Queried ChatGPT with mathematics and figures to assess alt-text viability.
\item \textit{Stage 2: PowerPoint Remediation Prototype (Mar.--Apr.):}
  Converted the feasibility probe into a command-line tool for PowerPoint: walk slide content and query OpenAI for figures and equations. Colleagues requested PDF support.

\item \textit{Stage 3: Format Expansion (April):}
  Generalized PowerPoint support to Word and Excel. 
  Prototype support for PDF.
  Documents from many sources exposed assumptions.

\item \textit{Stage 4: SaaS (April):}
  Added service capabilities for institutional use, \eg a website, Cloud deployment, and authentication. Feedback from users at multiple institutions exposed requirements that had not appeared locally. One user remediated his course's slide decks with minimal concerns. Users also reported cases where outputs passed accessibility checkers but lost semantic equivalence.

\item \textit{Stage 5: Standards Conformance (April--May):}
  Shifted validation from external syntax-style accessibility checkers to a custom, standards-grounded approach. Existing checkers ignore many semantic obligations of WCAG.

\item \textit{Stage 6: Hardening (May--June):}
  The DOJ extended the deadline by one year~\cite{doj2026adaComplianceExtension}.
  Project shifted from emergency deadline response to production hardening, incorporating more disciplined architecture and expanded tests.
  
\end{enumerate}

\subsubsection{Tools}
\label{sec:case:agentic-environment:tools}

As an AI tool, the Subject used Anthropic's Claude.
He used the ``Claude for VSCode'' plugin, interacting entirely through that chat interface.
The Subject was responsible for goals, acceptance criteria, architectural direction, validation, and deployment decisions.
Claude implemented.

The Subject inspected almost no agent-produced code. 
He deliberately departed from the oversight-centric model of agentic quality (\cref{sec:background:process-models}), under which high-velocity agentic development depends on continuous human inspection of agent-produced code.
The case tested whether quality could be sustained without ongoing inspection (cf.~\cref{fig:case-process-overview}). 

\subsubsection{Cost}
\label{sec:case:system-cost}

The development cost was $\sim$\$\TotalDevCostRounded USD:
  \$\SalaryCostRounded for the Subject's salary (paid by research grants),
  \$\InferenceCostRounded for inference (OpenAI API, Google Vertex),
  \$\HostingCostRounded for Google Cloud (SaaS hosting),
  and
  \$\ClaudeCostRounded for Claude subscriptions.\footnote{The Subject saturated between two and four Claude Max 20x accounts each week. Anthropic has opaque usage data~\cite{anthropic2026UsageLimitBestPractices}, but estimates~\cite{tbuckworth2026ClaudeCodeMax20xLimits} suggest this project consumed 9--18M tokens per week during the study period.} 

\subsubsection{System State}
\label{sec:case:system-state}
Currently, the system can process Office and PDF files to pass or substantially improve performance on accessibility checks.
Slide decks for the Subject's courses process in approximately 60 seconds, costing $\sim$\$1 per deck.

\section{Results}
\label{sec:results}

This section reports the empirical findings from the incident analysis.
We first characterize the coded incident corpus to explain the analytic focus of the results (\cref{sec:results:coding}).
We give examples of the recurring failure-to-governance pattern (\cref{sec:results:incidents}),
and describe the governance mechanisms that accumulated (\cref{sec:results:controls}).

\subsection{Incident-Level Coding}
\label{sec:results:coding}

Our unit of analysis is the critical incident:
  an episode of failure, decision, acceleration, or reframing that the Subject felt salient to capture. 
\Cref{tab:coding} reports their distribution by class and category.
Layer~1 (\textit{class}) separates incidents that yield
  a software-engineering insight (\emph{engineering-reflection})
  from
  those
  concerning people, institutions, and the engineer's experience (\emph{context-and-relational}),
  and project chronology (\emph{temporal-narration}).
Layer~2 (\textit{category)} names the locus of the insight.
For example, for engineering reflection, categories include controls, architecture, and economics.
For the primary class, engineering reflection,
  we subdivided categories into ``process'' and ``product''.
For \textit{control} incidents, we recorded the resources each control made more governable.
These incident codes anchored subsequent thematic analysis (\cref{sec:theory}).

The corpus is dominated by engineering-reflection (72 of the 88 incidents).
Its two largest categories name two ways the Subject made a recurring failure governable.
\emph{Controls} (35 incidents), \ie checks such as lints, tests, validators, gates, and mediators, \emph{detect} a failure once an agent commits it.
\emph{Architecture} (20 incidents), such as typed models, bounded seams, and closed vocabularies, \emph{eliminates} the failure class by construction, so an agent can no longer express it.\footnote{Some mechanisms involve both elements, and these were coded by their primary function. For example, a typed model with lint support for violations.}
Together these categories are the analytic center of gravity of the study:
  \cref{sec:results:incidents} illustrates both modes, one architecture episode and one control episode,
  and
  \cref{sec:results:controls} characterizes the resulting repertoire, which spans construction and detection.




\begin{table}[t]
\caption{Coded incident corpus. The two largest engineering-reflection categories, Controls and Architecture, are the detection and construction responses to recurring failure.}
\centering
\label{tab:coding}
\begin{tabular*}{0.9\columnwidth}{@{\extracolsep{\fill}}llc@{}}
\toprule
\textbf{Class} & \textbf{Category} & \textbf{Incidents} \\
\midrule
\multirow{5}{*}{\textit{Engineering reflection (72)}}
 & Controls           & 35 \\
 & Architecture       & 20 \\
 & Economics          & 8  \\
 & Conceptual framing & 7  \\
 & AI capability      & 2  \\
\addlinespace
\multirow{3}{*}{\textit{Context and relational (10)}}
 & Personal affect         & 7 \\
 & External engagement     & 2 \\
 & Institutional \& market & 1 \\
\addlinespace
\multirow{3}{*}{\textit{Temporal narration (6)}}
 & Milestone & 3 \\
 & Origin    & 2 \\
 & Staffing  & 1 \\
\bottomrule
\end{tabular*}
\end{table}

\subsection{Examples of Architecture and Control Incidents}
\label{sec:results:incidents}

Most engineering-reflection episodes show a repeated pattern.
The Subject interpreted failures as structural problems rather than isolated defects, and made each failure class governable.
He either changed the architecture so the failure class could no longer arise, or added a control to detect it.
We illustrate both with two linked incidents: the first eliminates a failure class by construction (architecture), the second detects failures earlier with a control.

\subsubsection{Architecture: Modeling the system so enforcement can scale}
\label{sec:results:incidents:ComponentModel}

The recurring failure class was architectural enforcement that did not scale with the codebase.
The Subject initially relied on agents to check agent work,
  encoding the system's structural invariants (\eg which components may cross which boundaries) as agent briefs and running an ``agent audit crew'' to validate those properties.
As the codebase grew, a whole-repository audit no longer fit an agent's context, and audits were scoped by architectural ``zone.''
The audits were expensive and non-deterministic, and structural drift accumulated.

The resource at risk was architectural integrity:
  component boundaries and cross-cutting invariants eroded, and the enforcement mechanism grew more costly as the system it governed grew.
The Subject interpreted this as a modeling problem, in the sense of model-based software engineering (MBSE).
The architecture itself existed, but had no durable representation.  
The Subject therefore reified the audit zones through the mechanism of a \textit{typed component catalog}, in which each zone became an object naming its filesystem scope, its boundary kind, and relevant checks.
This architectural change converted agent-mediated audit into deterministic enforcement.
At first, the component model scoped the audit crew's sweeps.
Then, zone by zone, the audit crew was replaced by static and dynamic analyses using the model.

\subsubsection{Control: Shifting left on failures via dynamic context injection}
\label{sec:results:incidents:DynamicContextInjection}

The recurring failure class was constraint under-specification at agent dispatch time.
A coding agent lacked the Subject's tacit knowledge of which lints, conventions, tests, and component boundaries applied to a target change.
Agents therefore made plausible edits that violated those constraints.
This created a throughput problem: 
  agents eventually resolved their errors but wasted time and tokens doing so.
The existence of a validation hierarchy (lightweight git precommit, medium merge check, strong deployment gates) exacerbated the problem, with context loss between authoring and repair.
\begin{table*}[t]
\centering
\caption{Catalog of governance mechanisms developed in the case, organized by target and family (cf.~\cref{fig:case-process-overview})}
\label{tab:control-catalog}
\begin{tabular}{p{0.180\linewidth}p{0.34\linewidth}p{0.40\linewidth}}
\toprule
\textbf{Family} & \textbf{Definition} & \textbf{Representative mechanisms} \\
\midrule
\addlinespace[1pt]
\multicolumn{3}{l}{\ul{\textit{Agent governance mechanisms constrain agent behaviors}}} \\
\addlinespace[3pt]
Governance-doc controls &
Mechanisms that make agent-facing project rules enforceable rather than conventional. &
\texttt{CLAUDE.md} rule index and cap lint; mandatory snippet-table enforcement; Epic templates and Definition-of-Done checks. \\

Context \& dispatch &
Mechanisms that shape what an agent knows, what role it receives, and under what constraints. &
Brief-linting; dynamic context injection; role-typed dispatch; hierarchical documentation and indices. \\

Agent observability &
Mechanisms that emit live state about agents, workers, deploys, and substrate health. &
Agent registry; typed event bus and playbook; sentinel and tombstone commits; cron-alerts gate. \\

Resource mediators &
Mechanisms that serialize or ration shared host resources across concurrent worktrees. &
Test serializer; build serializer; agent admission protocol based on machine health and agent task. \\

Incorporation gates &
Staged incorporation mechanisms for agent work --- progressively stronger checks. &
Sentinel first-commit early abort; pre-commit hook; merge-train batching; staged deploy gates. \\

\midrule
\addlinespace[1pt]
\multicolumn{3}{l}{\ul{\textit{Product governance mechanisms promote product quality}}} \\
\addlinespace[3pt]
Canonical seams &
Sanctioned models and boundaries for product structure and cross-boundary interaction. &
Per-document models and iteration primitives. \\

Validation and conformance &
Deterministic checks over accessibility obligations and document integrity. &
Standards/WCAG rule engine; file corruption checks. \\

Static \& dynamic analyses &
Analyses that enforce software invariants or pin expected behavior. &
Custom static analyses; unit/function/e2e tests; property-based testing; fuzzing. \\

Provenance \& attribution &
Mechanisms that record what changed, why, and through which mutation surface. &
Per-mutator attribution stamps; Changelog. \\ 

Repair vocabulary &
Mechanisms that bound remediation actions and failure categories to closed, typed sets. &
Closed remediation-verb sets; codemod-first threshold for lint repairs. \\
\bottomrule
\end{tabular}
\end{table*}

The resource at risk was productive agent throughput: the time and tokens an agent spent discovering and repairing constraint violations it could have been told to avoid.
The relevant context already existed in the repository, but it was distributed across component models, lint declarations, repair documentation, and project conventions.
The engineering problem was to make that context selectable and inject it before work began.
The Subject installed a new mechanism: the \textit{dynamic context injection} control.
Building on the repository's system models, the Subject designed a dispatch-time constraint slicing mechanism to map a change's intended files to the precise constraints that govern them, and inserted that subset into the agent's task brief before the agent edited code.
This moved detection ``left'' of the cheapest traditional gate: rather than waiting for an agent to violate a lint and then repair the failure, the orchestrator briefed the agent on applicable rules.


\subsection{The Governed Engineering Environment}
\label{sec:results:controls}

As we saw in the example incidents~\cref{sec:results:incidents}, the Subject repeatedly introduced mechanisms to provide governance.
We catalogued ten distinct families of governance mechanisms from the incidents and repository records (\cref{tab:control-catalog}).
To do this, we used the incident data to identify candidate mechanisms, then analyzed the repository's governance substrate (\eg the agent-legible system models and the static-analysis registry) to corroborate and categorize them.
The same legibility that makes these mechanisms usable by agents also made it tractable to identify and categorize them for analysis.

Many mechanisms were layered.
When a failure recurred across one mechanism, the Subject's response was to add a complement.
The deepest stack, 14 complementary mechanisms, accreted around high-frequency failure classes in Orchestrator/Sub-agent interactions. 
These defenses were tailored to the nature of the agents' errors, and produced a belt-and-suspenders governed engineering environment.

The governed environment is large relative to the product.
The catalog records \NumControlsCatalog representative mechanisms across \NumControlFamilies families. 
Measured by source size, the support apparatus comprises \SupportApparatusMLOC, or \SupportRatio the production code ---
   static and dynamic analyses (\LintsKLOC and \TestsKLOC, respectively),
   agent-referenced documentation (\LoadBearingDocsKLOC),
   agent infrastructure (\AgentInfraKLOC),
   and tooling (\ToolsKLOC).
The static-analysis controls include custom and standard checks: \NumCustomLints project-specific analyses supported by commodity analyzers provided through community rules for Roslyn/.NET, Pyright/Python, and ESLint/TypeScript. 
For dynamic analysis, there are over \NumCSharpTestMethods C\# test methods and roughly \NumPythonTestFiles Python test files, including more than \NumPropertyTests property-based tests and \NumFuzzHarnesses fuzz harnesses.
These mechanisms enforce quality without the Subject's implementation-level oversight.

\begin{figure*}[t]
\centering
\scriptsize
\setlength{\tabcolsep}{2.5pt}
\renewcommand{\arraystretch}{1.06}

\begin{minipage}[t]{0.48\textwidth}

\begin{subfigure}[t]{\linewidth}
\centering
\caption{Proposed process}
\label{fig:theory:mechanism}
\begin{tabular}{@{}p{0.38\linewidth}p{0.60\linewidth}@{}}
\toprule
\textbf{Process step} & \textbf{Explanation} \\
\midrule
1. Velocity exposes failure
&
Agent-produced changes surface ambiguity, drift, weak boundaries, and missing oracles quickly. \\
\addlinespace

2. Human classifies failure
&
The architect distinguishes local defects from recurring structural failure classes. \\
\addlinespace

3. Governance conversion
&
Governance is encoded probabilistically (\eg agent harness) or deterministically (\eg type system). \\
\addlinespace

4. Future work is constrained
&
Subsequent agents inherit a narrower and more explicit action space. \\
\addlinespace

5. Governability compounds
&
Repeated conversions increase the environment's capacity to absorb future agent work. \\
\bottomrule
\end{tabular}
\end{subfigure}

\vspace{0.8em}

\begin{subfigure}[t]{\linewidth}
\centering
\caption{Constructs and scope conditions}
\label{fig:theory:constructs}
\begin{tabular}{@{}p{0.40\linewidth}p{0.58\linewidth}@{}}
\toprule
\textbf{Condition / Construct} & \textbf{Working meaning} \\
\midrule
Scope condition
&
Sustained agentic work on a software system whose quality matters. \\
\addlinespace

Agentic implementation velocity
&
Rate at which agents generate, modify, repair, test, and document software under human direction. \\
\addlinespace

Human architectural-contextual judgment
&
Capacity to frame goals and govern agent work, \eg by interpreting failures. \\
\addlinespace

Structural failure class
&
A failure interpreted as a weak or absent abstraction or guardrail. \\
\addlinespace

Engineering governance substrate
&
The harnessing, types, lints, schemas, deployment processes, etc. necessary for agent productivity. \\
\addlinespace

\bottomrule
\end{tabular}
\end{subfigure}

\end{minipage}%
\hspace{0.015\textwidth}{\color{black!45}\vrule width 0.3pt}\hspace{0.015\textwidth}%
\begin{minipage}[t]{0.48\textwidth}
\begin{subfigure}[t]{\linewidth}
\centering
\caption{Propositions and moderators}
\label{fig:theory:propositions}
\begin{tabular}{@{}p{0.36\linewidth}p{0.60\linewidth}@{}}
\toprule
\textbf{Proposition / Moderator} & \textbf{Statement} \\
\midrule
Prop.: Velocity-exposure 
&
Higher agentic velocity surfaces structural failure classes more quickly. \\
\addlinespace

Prop.: Soft-control saturation 
&
Tacit conventions and review-centered controls saturate under high agentic velocity. \\
\addlinespace

Prop.: Governance-conversion 
&
Sustainable progress depends on converting recurring failures into explicit governance. \\
\addlinespace

\midrule

Moderator: Capability-fit 
&
Outcomes depend on fit between agent capability and human judgment. \\
\addlinespace

Moderator: Authority 
&
Engineers must be able to alter architecture and controls, lest structural failures get local patches. \\
\bottomrule
\end{tabular}
\end{subfigure}

\vspace{0.8em}

\begin{subfigure}[t]{\linewidth}
\centering
\caption{Empirical indicators and outcomes}
\label{fig:theory:indicators}
\begin{tabular}{@{}p{0.38\linewidth}p{0.60\linewidth}@{}}
\toprule
\textbf{Indicator / Outcome} & \textbf{Expected observation} \\
\midrule

Failure-to-governance episodes
&
Repeated pattern of observed failure, human interpretation, and durable governance installation. \\
\addlinespace

Control and architecture accumulation
&
Growth of governance substrate over time (\eg lints, validators, typed representations, and audit trails). \\
\addlinespace

Governable acceleration
&
Governance conversion compounds, allowing agentic throughput to continue or increase. \\ 
\addlinespace


Ungoverned churn
&
If controls or judgment are weak, agents produce plausible changes but not coherent software. \\
\addlinespace

Capability amplification
&
More capable agents and humans increase ambition and concurrency, amplifying the need for controls. \\

\bottomrule
\end{tabular}
\end{subfigure}
\end{minipage}

\caption{
  Proposed middle-range theory~\cite{merton1968social} 
  explaining how AI-mediated implementation becomes governable software engineering progress.
  Briefly:
    agentic velocity exposes structural failure classes
    $\rightarrow$
    human architectural-contextual judgment interprets those failures
    $\rightarrow$
    structural failures are converted into explicit governance constraining subsequent agent work.
  }
\label{fig:theory}
\end{figure*}

\section{A Middle-Range Theory of Governance Conversion in Agentic Software Engineering}
\label{sec:theory}

This section synthesizes the case into a candidate middle-range process theory of \emph{governance conversion}: a process by which sustained agentic velocity becomes governable software-engineering progress.

\subsection{Middle-Range Theory}
\label{sec:theory:mrt}

We frame our synthesis as a \emph{middle-range theory}~\cite{merton1968social}, \ie
  an explanation somewhere in the ``middle'' between a working hypothesis and an all-encompassing general theory of software engineering.
Much engineering knowledge has this character.
Brooks's Law~\cite{brooks1995mythical}, Conway's Law~\cite{conway1968committees}, Hyrum's Law~\cite{winters2020software}, and Cunningham's technical-debt metaphor~\cite{cunningham2009debtmetaphor} all emerged from expert interpretation of situated experience~\cite{schon2017reflective}.
We do not assert prevalence from one case.
We propose a process with its constructs and relationships so that researchers can test its boundary conditions (\cref{sec:theory:predictions}), tool builders can design for it, and practitioners can be trained in it (\cref{sec:discussion}).

\subsection{The Theory: Cheap Code, Costly Judgment}
\label{sec:theory:theory}

\subsubsection{Empirical Grounding}

This theory is the engineering framing of our thematic analysis of the corpus.
We identified thirteen cross-incident themes, grouped into four clusters.
Three clusters directly ground the theory:
  \textit{abundant implementation but scarce judgment} grounds the human judgment thesis;
  \textit{governance relocated to the substrate} grounds the engineering governance thesis and conversion loop;
  and
  \textit{engineering an agent-legible system} grounds the governance-substrate construct and capability-fit moderator.
The fourth cluster,
  \textit{the reflexive build and the evolving engineer}, characterizes the study's first-person method.
That cluster informed our Method rather than the theory. 

\subsubsection{Theory}
\Cref{fig:theory} maps the theory we induced:
  the process 
  (\cref{fig:theory:mechanism}),
  constructs and scope (\cref{fig:theory:constructs}),
  propositions and moderators (\cref{fig:theory:propositions}),
  and
  indicators and outcomes (\cref{fig:theory:indicators}).
We do not claim it is universal, but we assert it captures a distinct and useful perspective.

\subsubsection{The Cheap-Code, Costly-Judgment Loop}

As implementation becomes abundant, judgment becomes relatively scarce.
The relevant judgment is not only whether an agent-produced change is acceptable, but whether a failure reveals missing governance.
Engineers must decide what and how to build, interpret failures, and design the governance under which fast code can be trusted.
The process is a loop (\cref{fig:theory:mechanism}, cf.~\cref{fig:case-process-overview}): velocity exposes a failure class, the architect classifies it as local or structural, structural failures convert into governance, and subsequent agents inherit a narrower action space.
The loop is non-terminating, as increasing velocity surfaces failure classes prior governance did not address.
The governed engineering environment emerges iteratively, shaped by joint human-agent capability.

\subsubsection{Two Dual Theses}

The loop yields two linked theses.

\textbf{(a) The engineering governance thesis}:
Agentic velocity is possible, but is sustainable only when governance is relocated into the engineering environment
rather than resting on human review. 
Under agentic velocity governance must become \emph{machine-actionable} enough to shape agent work lest human review become the bottleneck.

In the case, we observed that weak forms of governance fail with project and agent scale. 
The Subject's environment does encode ``soft'' probabilistic controls, such as templates that an Orchestrator should follow when preparing agent briefs and work Epics.
But it was necessary to mate these to deterministic controls, such as
  typed enumerations the compiler checks
  and
  static and dynamic analyses gating commits.
At velocity, a low-probability agent harness violation becomes a certainty.
Perhaps counterintuitively, stronger controls do not always tax agentic velocity; they can compound it. 

\textbf{(b) The human judgment thesis}: 
When implementation becomes abundant, human work shifts toward problem framing, abstraction discovery, architectural judgment, and governance~\cite{thawar2026whatjobnow}.
Better agents make this work more consequential, not less, because these decisions determine whether throughput becomes durable progress or churn.
On this, the Subject wrote: ``\textit{Now I feel like an engineer, and more than an engineer}.''

Governance does not design itself.
Each element required Subject to distinguish local defects from structural failures and propose an intervention. 
Two moderators bound this thesis (\cref{fig:theory:propositions}): \emph{capability-fit}, as agents must be matched with suitable human judgment, and \emph{authority}, as the human must be able to alter the engineering environment. 

\vspace{4pt}
\noindent\hspace*{0.5\parindent}\textbf{Synthesis.}\enspace Together, the theses are duals: governance makes velocity sustainable, and judgment determines which governance should exist.
Agents' current implementation capabilities do not supply the structural interpretation needed to convert recurring failures into durable governance.

\subsection{Testable Propositions from the Theory}
\label{sec:theory:predictions}

Because this study is theory-building rather than theory-testing, we describe candidate constructs and relationships but do not estimate their prevalence or effect sizes.
To make the theory empirically accountable, we articulate falsifiable propositions for future work~\cite{popper2012science} (\cref{fig:theory:indicators}), including:

\textit{Does velocity expose structure?}
Higher agentic velocity should surface a structural failure class in fewer calendar days than lower-velocity development.
Repository mining across projects of varying throughput could measure the lag from a class's first instance to its structural fix.
The case grounds the premise but cannot parameterize velocity models. 

\textit{Do soft controls saturate?}
As agent velocity increases, review-based quality regimes should show rising escaped-defect rates while control-based regimes do not.
Controlled experiments could establish this relationship. 

\textit{Does conversion compound?}
Teams that convert failures into deterministic controls should sustain velocity, and teams relying on inspection should not, as work repeats failures.
A longitudinal comparison could test this, though note that we found compounding to be a tendency, not strict monotonicity. 

\textit{Do judgment and authority shape outcomes?}
Holding agent capability fixed, outcomes should depend on human judgment and on the authority to modify the environment, with structural failures persisting where engineers can only request local patches.
This prediction might be most amenable to human factors methods such as interviews and surveys. 

\vspace{2pt}
\noindent\hspace*{0.5\parindent}\textbf{Scope:}\enspace
These propositions pertain to agentic work on systems where correctness and maintainability matter (\cref{fig:theory:constructs}). 

\section{Discussion}
\label{sec:discussion}

\subsection{What Does our Theory of Governance Conversion Add?}
\label{sec:discussion:ExPost}

\subsubsection{Beyond ex-ante governance}
Our study provides a worked example of governance-centric agentic software engineering, but also revises that model.
Existing governance-centric accounts emphasize \emph{ex-ante} governance: known obligations are identified in advance and translated into controls before agents act (\cref{sec:background:process-models}).
The Subject did use \textit{ex-ante} governance,
  such as
  WCAG~2.1~AA conformance checkers (\MyInlineRequirement{2}),
  content preservation and provenance for auditability (\MyInlineRequirement{3}),
  and
  commodity lints (\eg pyright, ESLint).
We found these controls necessary but not sufficient.

\subsubsection{Failure as discovery}
The theoretical delta is not simply that agent work needs
controls.
All approaches agree on that~(\cref{sec:background}).
The new claim is that, under agentic velocity, many necessary controls cannot be fully specified before work begins.
The necessary controls are discovered through failure: agents make plausible but wrong changes.
To sustain velocity, an engineer must interpret those changes as ambiguities, missing abstractions, underspecified interfaces, and coordination gaps, and convert them into new architecture and controls.
Our theory of governance conversion extends governance-centric accounts from known-obligation enforcement to failure-induced control discovery.

\subsubsection{Contrast with the Velocity and Oversight Models (\cref{sec:background:process-models})}
The velocity-centric view provides agents with soft guardrails.
Our case suggests that without a stronger governed engineering environment, agent velocity will not yield progress.
The oversight-centric view recommends that humans supervise agent changes.
The Subject instead analyzed agent \textit{failures}.
Work that produced no failure signal was admitted; work that exposed a recurring failure class led to new governance.
Our contribution is to describe this process mechanism: at velocity, engineering leverage comes from converting a failure mode once so that future instances of its class require less oversight.

\subsection{Software Engineering is Old and New Again}
\label{sec:discussion:SWEBOK}

\subsubsection{Repertoire, not Recipe}
The Subject's high-velocity AI-mediated implementation depended on familiar methods, including requirements analysis, architecture, validation, infrastructure-as-code, documentation, and process control.
What changed was where those methods had to reside.
In conventional development, engineering discipline can remain tacit or social~\cite{bjornson2008knowledge,anandayuvaraj2026learning}.
Under agentic velocity, the Subject's engineering judgment became effective only when externalized into an agent-legible and machine-actionable governance substrate.
In this sense, ``conventional engineering'' was enough, but only when converted into durable mechanisms that constrained subsequent agent work.

But if classical SE is enough, why not simply encode it into the \textit{ex-ante} governance model?
We suggest that both \textit{ex-ante} and \textit{ex-post} governance are needed.
The Subject applied the lessons from this case as \textit{ex-ante} governance on two subsequent projects (50\,KLOC and 36\,KLOC), obtaining faster initial progress.
Even so, \textit{ex-post} governance remained necessary: new failure modes appeared as each project and its development substrate evolved.
The precise governance required therefore depends on the product, its trajectory, and the changing failure modes of the agents that act on it.

\subsubsection{Agents Change the Marginal Cost of Quality}
Agentic velocity reduced the cost of refactoring and architectural improvement, making quality-improving changes feasible at scales that would normally be deferred.
In one incident, over two days the Subject refactored an entire web front-end to TypeScript ($\sim$100 KLOC), with no defects observed under the project’s checks; and induced a type system supported by a lint for excessive primitive density.
In another incident, agents surfaced related failures in dense succession, which would
conventionally surface far apart in time. 
Their proximity allowed the Subject to perceive an architectural gap.
He introduced a new system model enabling new static analyses, closing several classes of failures.
\textit{The same speed that can cause agentic technical debt~\cite{EhsaniDebt2026,agarwal2026ai} can also make quality-improving changes more affordable and visible.}

\subsection{Implications for Software Engineering}
\label{sec:discussion:OrgQuestions}

\subsubsection{Governance Requires Authority}
Agentic tools compress team-scale coordination problems into individual workflows.
In the case, a single human architect produced enough change through agents to require artifacts normally associated with team coordination, such as architectural models, task contracts, and staged incorporation.
This compression makes authority central.
Governance conversion requires the person who observes recurring failure to alter the engineering environment.
In organizations, where authority may be divided across teams, owners, review boards, and deployment processes~\cite{conway1968committees}, the same failure signal may produce only local patches unless there is an organizational path for converting it into shared governance.

\subsubsection{Productivity Metrics}
Our case suggests that the relevant productivity metric is \textit{governed throughput}, not implementation volume (\eg ``tokenmaxing'').
Commits landed, lines changed, and agent tasks completed measure activity, but not whether that activity becomes durable progress.
Organizations should therefore ask whether agent-produced failures are converted into reusable governance, and whether later work benefits from those conversions without degrading coherence.

\subsection{GenAI Paradox: Agentic Velocity by Reduced Reasoning}
\label{sec:discussion:paradox}

Generative models are valuable because they operate over open-ended inputs, yet Subject's method was to use governance to systematically reduce the opportunities for unconstrained reasoning. 
His method can be understood as a response to the long-horizon problem~\cite{kwa2026measuring}.
Software agents perform complex tasks~\cite{jimenez2024swe}, with a per-step probability of error~\cite{sinha2025illusion}.
The Subject's response was not to exclude probabilistic model behavior, but to bound it:
  agents received soft guidance in their briefs, faced deterministic guardrails on their actions, and their outputs were admitted only through a deterministic control envelope.

A corollary concerns how engineers should read model failure.
When a delegated task fails, the tempting inference is that the model is incapable of it.
The case repeatedly suggests another interpretation: the task may not yet have been framed as an inspectable, controllable process.
The Subject's approach was, in one incident, to ``\textit{assume that you are the problem}''.
We advance this as an engineering disposition: engineers must relocate the locus of difficulty from model competence to their own capacity to frame, decompose, and constrain.

\subsection{Implications for Research and Education}
\label{sec:discussion:Implications}

The case demonstrates that current models change the cost calculus of software engineering.
In the Subject's expert hands, agents produced a deployed product at agentic speed.
The research community should attend not only to raw agent capabilities on benchmarks, but also to changes in the surrounding software engineering process~\cite{gao2025SELLMChallenges,sarkar2025vibecoding}.
Future research can treat human-agent interaction and the engineering substrate as first-class parts of the process.
We know agents can generate.
What engineering environments make that generation governable?
Can agents discern inadequate governance, and would this improve with better architectural awareness~\cite{peng2026beyond}?\footnote{In the case, Claude did not initially offer governance responses, but began suggesting plausible interventions as the codebase accumulated examples.}
What tools help engineers distinguish local from structural defects?

As software agents commoditize implementation, our study points to a shift in the software engineer's role toward architectural and contextual judgment \cite{sarkar2025vibecoding}.
The Subject provided abstraction design, requirements clarification, validation, and process evolution.
He inspected little agent-produced code, but his supervision depended on accumulated software-engineering knowledge about design, architecture, validation, and process.
This interpretation is consistent with emerging discussions in software engineering education and workforce development, which similarly emphasize specification, validation, and AI use over syntax-level implementation~\cite{kirova2024SEShouldAdaptLLM,randall2024AISECurriculum}.

\section{Threats to Validity}
\label{sec:threats}

\subsection{Construct Validity}
\label{sec:construct-validity}

Our primary constructs were agentic governance and velocity.
Governance is an emerging paradigm with under-determined constructs.
We interpreted but did not measure this construct, rather offering an alternative view on the constructs and mechanisms that underpin it.
For velocity, we reported conventional software metrics (\cref{fig:repo-activity}), but these are hard to interpret.
Code is only a proxy for complexity, and documentation is a probabilistically executable agent constraint, blurring typical distinctions between code and documentation.


\subsection{Internal Validity}
\label{sec:internal-validity}

This study examines a single Subject (\textit{N=1}) using one agentic engineering toolchain.
The resulting observations may reflect characteristics of the subject and the particular toolchain employed rather than universal properties of agentic software engineering.
In addition, the Subject participated in the retrospective interpretation of development artifacts, introducing the possibility of selective memory and confirmation bias.
Interpretive factors are mitigated by multi-author review (\cref{sec:method}).

A first-person method is both a strength and a limitation.
It provides direct access to experiences and rationales that are invisible in other methods.
It also risks selective memory, retrospective rationalization, and overinterpretation.
We mitigated these risks by grounding the analysis in contemporaneous field notes, repository evidence, and with a critical author team. 

\subsection{External Validity}
\label{sec:external-validity}

As discussed in \cref{sec:Method:Participant}, the Subject possessed substantial prior software engineering experience. Whether comparable outcomes could be achieved by non-experts, domain specialists, or larger teams remains an open question.
Future advances in model capabilities may alter the expertise requirements and coordination patterns identified here.
Our theory deliberately frames the outcomes as mediated by human-agent capability fit, which should insulate it from incremental improvements in agent performance but not from substantial ones.
This case study should be interpreted as characterizing contemporary agentic software engineering.


The case involved a one-subject development process, not a conventional software team. 
Large organizations introduce additional coordination problems, including ownership boundaries, divided authority and decision-making hierarchies, and regulatory requirements, handoffs, and review obligations.
We therefore expect the mechanism of governance conversion to remain relevant in teams, but its realization would likely require organizational processes for assigning authority, resolving conflicting judgments, and incorporating new controls into shared engineering infrastructure.

\section{Research Ethics}
\label{sec:Ethics}

This study involved self-experimentation by an author (Subject).
Sustained agentic development was harmful:
  obsessive tool use caused
  degraded attention to personal relationships
  and physical strain from prolonged work sessions.
High-velocity agentic tools make extreme work patterns practically sustainable for longer than is healthy.

\ifARXIV
\section{Open Science}
\label{sec:openscience}

A detailed description of the governed engineering environment, including an enumeration of the controls as well as accompanying infrastructure and CLAUDE skill, is available at \url{https://davisjam.github.io/agent-governance-mechanisms/index.html}.

\fi

\section{Conclusion}
\label{sec:conclusion}

This case study provides a view on the changing nature of software engineering in the agentic era.
In industry and the academic literature, it is unclear whether and how agentic velocity can be governably sustained.
Our study suggests an answer:
  \textit{Agentic velocity can be sustained when agents operate within a governed engineering environment, and the achievable velocity depends on the maturity of the governance.}
We present a candidate middle-range theory articulating the relevant constructs and mechanisms.
Our theory is \textit{cheap code, costly judgment} --- the future of software engineering lies in judgment that informs product direction and becomes embodied in the architecture and controls necessary for agents to make progress.
The research community is at an early stage of characterizing agentic software engineering.
Public, long-running accounts of the process by which human judgment and AI capability interact in real engineering work are the basis from which mature theory can be built.

\ifARXIV
\section*{Acknowledgments}
We thank
  H. Peng, R. Calvo, and L. Parente, for their thoughtful comments.
We thank
  Purdue College of Engineering and the American Society for Engineering Education (ASEE'26) for opportunities to discuss our work.
This work was funded by the US National Science Foundation under awards 
  \#2541917 and \#2452533.
\fi

\JD{I have introduced some references from GPT without checking them (checking 2-3 out of a batch of 5, for example). We need to check each reference for existence.}

\clearpage

\bibliographystyle{IEEEtran}
\bibliography{bib/bib}

@book{merton1968social,
  author    = {Robert K. Merton},
  title     = {Social Theory and Social Structure},
  edition   = {1968 Enlarged Edition},
  publisher = {Free Press},
  address   = {New York, NY, USA},
  year      = {1968}
}

@book{brooks1995mythical,
  author    = {Frederick P. Brooks, Jr.},
  title     = {The Mythical Man-Month: Essays on Software Engineering},
  edition   = {Anniversary Edition},
  publisher = {Addison-Wesley},
  address   = {Reading, MA, USA},
  year      = {1995}

}

@article{conway1968committees,
  author  = {Melvin E. Conway},
  title   = {How Do Committees Invent?},
  journal = {Datamation},
  volume  = {14},
  number  = {4},
  IGNOREpages   = {28--31},
  year    = {1968}
}

@book{winters2020software,
  title={Software engineering at google: Lessons learned from programming over time},
  author={Winters, Titus and Manshreck, Tom and Wright, Hyrum},
  year={2020},
  publisher={" O'Reilly Media, Inc."}
}

@misc{cunningham2009debtmetaphor,
  author       = {Ward Cunningham},
  title        = {Debt Metaphor},
  howpublished = {YouTube video},
  year         = {2009},
  month        = feb,
  day          = {14},
  IGNOREurl          = {https://www.youtube.com/watch?v=pqeJFYwnkjE},
  note         = {Explains the technical debt metaphor}

}

@book{schon2017reflective,
  title={The reflective practitioner: How professionals think in action},
  author={Sch{\"o}n, Donald A},
  year={2017},
  publisher={Routledge}
}

@misc{amodei2026policy,
  author       = {Amodei, Dario},
  title        = {Policy on the {AI} Exponential},
  year         = {2026},
  month        = jun,
  url = {https://darioamodei.com/post/policy-on-the-ai-exponential},
  IGNOREnote         = {Accessed: 2026-06-12}
}

@inproceedings{jimenez2024swe,
  title={{SWE}-bench: Can Language Models Resolve Real-world Github Issues?},
  author={Carlos E Jimenez and John Yang and Alexander Wettig and Shunyu Yao and Kexin Pei and Ofir Press and Karthik R Narasimhan},
  booktitle={The Twelfth International Conference on Learning Representations},
  year={2024}
}

@article{he2025llm,
author = {He, Junda and Treude, Christoph and Lo, David},
title = {LLM-Based Multi-Agent Systems for Software Engineering: Literature Review, Vision, and the Road Ahead},
year = {2025},
issue_date = {June 2025},
publisher = {Association for Computing Machinery},
address = {New York, NY, USA},
volume = {34},
number = {5},
issn = {1049-331X},
doi = {10.1145/3712003},
journal = {ACM Trans. Softw. Eng. Methodol.},
month = may,
articleno = {124},
numpages = {30}
}

@article{dong2025survey,
  title={A Survey on Code Generation with LLM-based Agents}, 
  author={Yihong Dong and Xue Jiang and Jiaru Qian and Tian Wang and Kechi Zhang and Zhi Jin and Ge Li},
  year={2025},
  eprint={2508.00083},
  archivePrefix={arXiv},
  primaryClass={cs.SE},
  url={https://arxiv.org/abs/2508.00083},
}

@inproceedings{roy2024exploring,
author = {Roy, Devjeet and Zhang, Xuchao and Bhave, Rashi and Bansal, Chetan and Las-Casas, Pedro and Fonseca, Rodrigo and Rajmohan, Saravan},
title = {Exploring LLM-Based Agents for Root Cause Analysis},
year = {2024},
isbn = {9798400706585},
IGNOREpublisher = {Association for Computing Machinery},
IGNOREaddress = {New York, NY, USA},
IGNOREurl = {https://doi.org/10.1145/3663529.3663841},
doi = {10.1145/3663529.3663841},
booktitle = {Companion Proceedings of the 32nd ACM International Conference on the Foundations of Software Engineering},
IGNOREpages = {208–219},
numpages = {12},
location = {Porto de Galinhas, Brazil},
IGNOREseries = {FSE 2024}
}

@inproceedings{rondon2025evaluating,
author={Rondon, Pat and Wei, Renyao and Cambronero, José and Cito, Jürgen and Sun, Aaron and Sanyam, Siddhant and Tufano, Michele and Chandra, Satish},
  booktitle={2025 IEEE/ACM 47th International Conference on Software Engineering: Software Engineering in Practice (ICSE-SEIP)}, 
  title={Evaluating Agent-Based Program Repair at Google}, 
  year={2025},
  volume={},
  number={},
  IGNOREpages={365-376},
  doi={10.1109/ICSE-SEIP66354.2025.00038}
}

@misc{anthropic2026compiler,
  author = {Nicholas Carlini},
  title = {Building a C compiler with a team of parallel Claudes},
  year = {2026},
  month = {February},
  url = {https://www.anthropic.com/engineering/building-c-compiler},
}

@misc{cloudflare2026nextjs,
  author = {Steve Faulkner},
  title = {How we rebuilt Next.js with AI in one week},
  year = {2026},
  month = {February},
  url = {https://blog.cloudflare.com/vinext/}
}

@article{runeson2009guidelines,
author = {Runeson, Per and H\"{o}st, Martin},
title = {Guidelines for conducting and reporting case study research in software engineering},
year = {2009},
issue_date = {April     2009},
publisher = {Kluwer Academic Publishers},
address = {USA},
volume = {14},
number = {2},
issn = {1382-3256},
IGNOREurl = {https://doi.org/10.1007/s10664-008-9102-8},
IGNOREdoi = {10.1007/s10664-008-9102-8},
journal = {Empirical Softw. Engg.},
month = apr,
pages = {131–164},
numpages = {34}
}

@book{runeson2012case,
author = {Runeson, Per and Host, Martin and Rainer, Austen and Regnell, Bjorn},
title = {Case Study Research in Software Engineering: Guidelines and Examples},
year = {2012},
isbn = {1118104358},
publisher = {Wiley Publishing},
edition = {1st}
}

@article{peng2026sysllmatic,
title = {SysLLMatic: Large language models are software system optimizers},
journal = {Journal of Systems and Software},
volume = {240},
pages = {112929},
year = {2026},
issn = {0164-1212},
IGNOREdoi = {https://doi.org/10.1016/j.jss.2026.112929},
IGNOREurl = {https://www.sciencedirect.com/science/article/pii/S0164121226001627},
author = {Huiyun Peng and Arjun Gupte and Ryan Hasler and Nicholas John Eliopoulos and Chien-Chou Ho and Rishi Mantri and Leo Deng and Konstantin Läufer and George K. Thiruvathukal and James C. Davis}
}

@misc{yegge2026gastown,
  author       = {Yegge, Steve},
  title        = {Welcome to Gas Town},
  year         = {2026},
  month        = jan,
  day          = {1},
  url          = {https://steve-yegge.medium.com/welcome-to-gas-town-4f25ee16dd04},
  IGNOREnote         = {Accessed: 2026-06-13}
}

@inproceedings{bender2021parrots,
author = {Bender, Emily M. and Gebru, Timnit and McMillan-Major, Angelina and Shmitchell, Shmargaret},
title = {On the Dangers of Stochastic Parrots: Can Language Models Be Too Big?},
year = {2021},
isbn = {9781450383097},
IGNOREpublisher = {Association for Computing Machinery},
IGNOREaddress = {New York, NY, USA},
IGNOREurl = {https://doi.org/10.1145/3442188.3445922},
IGNOREdoi = {10.1145/3442188.3445922},
booktitle = {Proceedings of the 2021 ACM Conference on Fairness, Accountability, and Transparency},
IGNOREpages = {610–623},
numpages = {14},
location = {Virtual Event, Canada},
IGNOREseries = {FAccT '21}
}

@inproceedings{yangWhatPromptsDont2026,
title = "What Prompts Don{'}t Say: Understanding and Managing Underspecification in {LLM} Prompts",
    author = "Yang, Chenyang  and
      Shi, Yike  and
      Ma, Qianou  and
      Liu, Michael Xieyang  and
      Kaestner, Christian  and
      Wu, Tongshuang",
    IGNOREeditor = "Liakata, Maria  and
      Moreira, Viviane P.  and
      Zhang, Jiajun  and
      Jurgens, David",
    booktitle = "Findings of the ACL: {ACL} 2026",
    month = jul,
    year = "2026",
    IGNOREaddress = "San Diego, California, United States",
    IGNOREpublisher = "Association for Computational Linguistics",
    IGNOREurl = "https://aclanthology.org/2026.findings-acl.441/",
    pages = "9072--9101",
    ISBN = "979-8-89176-395-1"
}

@misc{yangAssessingImpactRequirement2026,
      title={Assessing the Impact of Requirement Ambiguity on LLM-based Function-Level Code Generation}, 
      author={Di Yang and Xinou Xie and Xiuwen Yang and Ming Hu and Yihao Huang and Yueling Zhang and Weikai Miao and Ting Su and Chengcheng Wan and Geguang Pu},
      year={2026},
      eprint={2604.21505},
      archivePrefix={arXiv},
      primaryClass={cs.SE},
      url={https://arxiv.org/abs/2604.21505}, 
}

@inproceedings{zhuoProSAAssessingUnderstanding2024,
title = "{P}ro{SA}: Assessing and Understanding the Prompt Sensitivity of {LLM}s",
    author = "Zhuo, Jingming  and
      Zhang, Songyang  and
      Fang, Xinyu  and
      Duan, Haodong  and
      Lin, Dahua  and
      Chen, Kai",
    IGNOREeditor = "Al-Onaizan, Yaser  and
      Bansal, Mohit  and
      Chen, Yun-Nung",
    booktitle = "Findings of the ACL: EMNLP 2024",
    month = nov,
    year = "2024",
    IGNOREaddress = "Miami, Florida, USA",
    IGNOREpublisher = "Association for Computational Linguistics",
    IGNOREurl = "https://aclanthology.org/2024.findings-emnlp.108/",
    doi = "10.18653/v1/2024.findings-emnlp.108",
    pages = "1950--1976"
}

@article{liuLostMiddleHow2024,
	title = "Lost in the Middle: How Language Models Use Long Contexts",
    author = "Liu, Nelson F.  and
      Lin, Kevin  and
      Hewitt, John  and
      Paranjape, Ashwin  and
      Bevilacqua, Michele  and
      Petroni, Fabio  and
      Liang, Percy",
    journal = "Transactions of the ACL",
    volume = "12",
    year = "2024",
    address = "Cambridge, MA",
    publisher = "MIT Press",
    IGNOREurl = "https://aclanthology.org/2024.tacl-1.9/",
    doi = "10.1162/tacl_a_00638",
    pages = "157--173"
}

@misc{mohsenimofidiContextEngineeringAI2026,
	title = {Context Engineering for {AI} Agents in Open-Source Software},
	url = {http://arxiv.org/abs/2510.21413},
	IGNOREdoi = {10.48550/arXiv.2510.21413},
	number = {{arXiv}:2510.21413},
	publisher = {{arXiv}},
	author = {Mohsenimofidi, Seyedmoein and Galster, Matthias and Treude, Christoph and Baltes, Sebastian},
	IGNOREurldate = {2026-06-15},
	date = {2026-02-05},
	eprinttype = {arxiv},
	eprint = {2510.21413 [cs.SE]},
}

@inproceedings{hongMetaGPTMetaProgramming2024,
title={Meta{GPT}: Meta Programming for A Multi-Agent Collaborative Framework},
author={Sirui Hong and Mingchen Zhuge and Jonathan Chen and Xiawu Zheng and Yuheng Cheng and Jinlin Wang and Ceyao Zhang and Zili Wang and Steven Ka Shing Yau and Zijuan Lin and Liyang Zhou and Chenyu Ran and Lingfeng Xiao and Chenglin Wu and J{\"u}rgen Schmidhuber},
booktitle={The Twelfth International Conference on Learning Representations},
year={2024},
IGNOREurl={https://openreview.net/forum?id=VtmBAGCN7o}
}

@inproceedings{qianChatDevCommunicativeAgents2024,
	title = "{C}hat{D}ev: Communicative Agents for Software Development",
    author = "Qian, Chen  and
      Liu, Wei  and
      Liu, Hongzhang  and
      Chen, Nuo  and
      Dang, Yufan  and
      Li, Jiahao  and
      Yang, Cheng  and
      Chen, Weize  and
      Su, Yusheng  and
      Cong, Xin  and
      Xu, Juyuan  and
      Li, Dahai  and
      Liu, Zhiyuan  and
      Sun, Maosong",
    IGNOREeditor = "Ku, Lun-Wei  and
      Martins, Andre  and
      Srikumar, Vivek",
    booktitle = "Proceedings of the 62nd Annual Meeting of the ACL (Volume 1: Long Papers)",
    month = aug,
    year = "2024",
    IGNOREaddress = "Bangkok, Thailand",
    IGNOREpublisher = "Association for Computational Linguistics",
    IGNOREurl = "https://aclanthology.org/2024.acl-long.810/",
    doi = "10.18653/v1/2024.acl-long.810",
    pages = "15174--15186",
}

@misc{chouBuildingSoftwareRolling2025,
	title = {Building Software by Rolling the Dice: A Qualitative Study of Vibe Coding},
	url = {http://arxiv.org/abs/2512.22418},
	IGNOREdoi = {10.48550/arXiv.2512.22418},
	shorttitle = {Building Software by Rolling the Dice},
	number = {{arXiv}:2512.22418},
	publisher = {{arXiv}},
	author = {Chou, Yi-Hung and Jiang, Boyuan and Chen, Yi Wen and Weng, Mingyue and Jackson, Victoria and Zimmermann, Thomas and Jones, James A.},
	IGNOREurldate = {2026-06-15},
    year = 2025,
	date = {2025-12-30},
	eprinttype = {arxiv},
	eprint = {2512.22418 [cs.SE]},
}

@misc{tangStudyDeveloperBehaviors2024,
	title = {A Study on Developer Behaviors for Validating and Repairing {LLM}-Generated Code Using Eye Tracking and {IDE} Actions},
	url = {http://arxiv.org/abs/2405.16081},
	IGNOREdoi = {10.48550/arXiv.2405.16081},
	number = {{arXiv}:2405.16081},
	publisher = {{arXiv}},
	author = {Tang, Ningzhi and Chen, Meng and Ning, Zheng and Bansal, Aakash and Huang, Yu and {McMillan}, Collin and Li, Toby Jia-Jun},
	urldate = {2026-06-15},
	date = {2024-05-25},
    year = 2024,
	eprinttype = {arxiv},
	eprint = {2405.16081 [cs.SE]},
}

@article{kaptein2026runtime,
      title={Runtime Governance for AI Agents: Policies on Paths}, 
      author={Maurits Kaptein and Vassilis-Javed Khan and Andriy Podstavnychy},
      year={2026},
      eprint={2603.16586},
      archivePrefix={arXiv},
      primaryClass={cs.AI},
      url={https://arxiv.org/abs/2603.16586}, 
}

@article{koch2026governance,
      title={From Governance Norms to Enforceable Controls: A Layered Translation Method for Runtime Guardrails in Agentic AI}, 
      author={Christopher Koch},
      year={2026},
      eprint={2604.05229},
      archivePrefix={arXiv},
      primaryClass={cs.AI},
      url={https://arxiv.org/abs/2604.05229}, 
}

@article{ait2025towards,
      title={Towards Automated Governance: A DSL for Human-Agent Collaboration in Software Projects}, 
      author={Adem Ait and Gwendal Jouneaux and Javier Luis Cánovas Izquierdo and Jordi Cabot},
      year={2025},
      eprint={2510.14465},
      archivePrefix={arXiv},
      primaryClass={cs.SE},
      url={https://arxiv.org/abs/2510.14465}, 
}

@article{kraprayoon2025ai,
title={AI Agent Governance: A Field Guide}, 
      author={Jam Kraprayoon and Zoe Williams and Rida Fayyaz},
      year={2025},
      eprint={2505.21808},
      archivePrefix={arXiv},
      primaryClass={cs.CY},
      url={https://arxiv.org/abs/2505.21808}, 
}

@article{rahimi2026agentic,
  title={Agentic and Multi-agent Systems: A Systematic Review of Tool Use, Benchmarks, and Governance},
  author={Rahimi, Audrey},
  year={2026}
}

@misc{anthropic2026claudepricing,
  author       = {{Anthropic}},
  title        = {{Plans \& Pricing | Claude by Anthropic}},
  year         = {2026},
  url = {https://claude.com/pricing},
  IGNOREnote         = {Accessed: 2026-06-16}
}

@misc{openai2026apipricing,
  author       = {{OpenAI}},
  title        = {{API Pricing}},
  year         = {2026},
  url = {https://openai.com/api/pricing/},
  IGNOREnote         = {Accessed: 2026-06-16}
}

@misc{ft2026amazon_ai_outages,
  author       = {{Financial Times}},
  title        = {{Amazon holds engineering meeting following AI-related outages}},
  url = {https://www.ft.com/content/7cab4ec7-4712-4137-b602-119a44f771de},
  month        = mar,
  year         = {2026},
  IGNOREnote         = {Published March 10, 2026; accessed June 17, 2026}
}

@misc{moonka2026amazon_sev1_ai,
  author       = {Moonka, Anish},
  title        = {{Amazon had four Sev-1 outages in a single week; internal memos say AI-assisted code changes were a contributing factor}},
  url = {https://x.com/anishmoonka/status/2031434445102989379},
  month        = mar,
  year         = {2026},
  IGNOREnote         = {X post; accessed June 17, 2026}
}

@misc{beck2001agilemanifesto,
  author       = {Beck, Kent and Beedle, Mike and van Bennekum, Arie and Cockburn, Alistair and Cunningham, Ward and Fowler, Martin and Grenning, James and Highsmith, Jim and Hunt, Andrew and Jeffries, Ron and Kern, Jon and Marick, Brian and Martin, Robert C. and Mellor, Steve and Schwaber, Ken and Sutherland, Jeff and Thomas, Dave},
  title        = {{Manifesto for Agile Software Development}},
  year         = {2001},
  url = {https://agilemanifesto.org/},
  IGNOREnote         = {Accessed June 17, 2026}
}

@inproceedings{royce1970managing,
  author    = {Royce, Winston W.},
  title     = {{Managing the Development of Large Software Systems: Concepts and Techniques}},
  booktitle = {Proceedings of IEEE WESCON},
  year      = {1970},
  IGNOREpages     = {1--9}
}

@book{saldana_coding_2013,
	address = {Los Angeles},
	edition = {2nd ed},
	title = {The coding manual for qualitative researchers},
	isbn = {978-1-4462-4736-5 978-1-4462-4737-2},
	language = {en},
	publisher = {Sage Publications, Inc},
	author = {Saldaña, Johnny},
	year = {2013},
	note = {OCLC: ocn796279115}
}

@misc{doj_titleii_finalrule,
  author       = {{U.S. Department of Justice}},
  title        = {{Nondiscrimination on the Basis of Disability; Accessibility of Web Information and Services of State and Local Government Entities}},
  year         = {2024},
  month        = apr,
  howpublished = {\emph{Federal Register}, 89 FR 31320--31396},
 IGNOREurl          = {https://www.federalregister.gov/documents/2024/04/24/2024-07758/nondiscrimination-on-the-basis-of-disability-accessibility-of-web-information-and-services-of-state},
  note         = {Final rule; Document No. 2024-07758; 28 CFR Part 35; effective June 24, 2024}
}

@misc{doj_effective_communication,
  author       = {{U.S. Department of Justice, Civil Rights Division}},
  title        = {{ADA Requirements: Effective Communication}},
  year         = {2020},
  month        = feb,
  url = {https://www.ada.gov/resources/effective-communication/},
  IGNOREnote         = {Last updated February 28, 2020; accessed May 13, 2026}
}

@misc{ada1990,
  author       = {{U.S. Department of Justice, Civil Rights Division}},
  title        = {{Americans with Disabilities Act of 1990, As Amended}},
  year         = {2026},
  url = {https://www.ada.gov/law-and-regs/ada/},
  note         = {ADA.gov; accessed June 17, 2026}
}

@misc{wcag21,
  author       = {{Accessibility Guidelines Working Group}},
  title        = {{Web Content Accessibility Guidelines (WCAG) 2.1}},
  year         = {2025},
  month        = may,
  IGNOREhowpublished = {{W3C Recommendation}},
  url          = {https://www.w3.org/TR/WCAG21/},
  IGNOREnote         = {W3C Recommendation, 06 May 2025; accessed June 17, 2026}
}

@inproceedings{kelly2021accessibility,
author = {Mack, Kelly and McDonnell, Emma and Jain, Dhruv and Lu Wang, Lucy and E. Froehlich, Jon and Findlater, Leah},
title = {What Do We Mean by “Accessibility Research”? A Literature Survey of Accessibility Papers in CHI and ASSETS from 1994 to 2019},
year = {2021},
isbn = {9781450380966},
IGNOREpublisher = {Association for Computing Machinery},
IGNOREaddress = {New York, NY, USA},
IGNOREurl = {https://doi.org/10.1145/3411764.3445412},
IGNOREdoi = {10.1145/3411764.3445412},
booktitle = {Proceedings of CHI},
articleno = {371},
numpages = {18},
location = {Yokohama, Japan},
}

@misc{verapdf_home,
  title        = {{veraPDF}: Industry Supported {PDF/A} Validation},
  author       = {{veraPDF Consortium}},
  year         = {2026},
  url = {https://verapdf.org/home/},
  IGNOREnote         = {Accessed: 2026-06-17}
}

@misc{adobe_acrobat_accessibility,
  title        = {Create and Verify {PDF} Accessibility},
  author       = {{Adobe}},
  year         = {2026},
  url = {https://helpx.adobe.com/acrobat/using/create-verify-pdf-accessibility.html},
  IGNOREnote         = {Accessed: 2026-06-17}
}

@misc{libreoffice_accessibility,
  title        = {Accessibility in {LibreOffice}},
  author       = {{The Document Foundation}},
  year         = {2026},
 url = {https://help.libreoffice.org/latest/en-US/text/shared/guide/accessibility.html},
  IGNOREnote         = {Accessed: 2026-06-17}
}

@misc{microsoft_accessibility_checker,
  title        = {Improve Accessibility with the Accessibility Checker},
  author       = {{Microsoft}},
  year         = {2026},
  url = {https://support.microsoft.com/en-us/accessibility/office-accessibility/improve-accessibility-with-the-accessibility-checker},
  IGNOREnote         = {Accessed: 2026-06-17}
}

@misc{doj2026adaComplianceExtension,
  author       = {{U.S. Department of Justice}},
  title        = {{Extension of Compliance Dates for Nondiscrimination on the Basis of Disability; Accessibility of Web Information and Services of State and Local Government Entities}},
  howpublished = {Federal Register, 91 FR 20902--20912},
  year         = {2026},
  month        = apr,
  day          = {20},
  IGNOREnote         = {Interim final rule; Docket No. CRT150; AG Order No. 6742-2026; RIN 1190-AA82; Document No. 2026-07663; effective April 20, 2026; comments close June 22, 2026; accessed June 19, 2026},
 IGNOREurl          = {https://www.federalregister.gov/documents/2026/04/20/2026-07663/extension-of-compliance-dates-for-nondiscrimination-on-the-basis-of-disability-accessibility-of-web}
}

@misc{tbuckworth2026ClaudeCodeMax20xLimits,
  author       = {{tbuckworth}},
  title        = {{Max 20x Plan Hitting Daily Limit with Reduced Usage---Limits Appear Silently Tightened (April 28--29, 2026)}},
  IGNOREhowpublished = {{GitHub issue \#54714, anthropics/claude-code}},
  year         = {2026},
  month        = apr,
  day          = {29},
  note         = {Closed as not planned; accessed June 19, 2026},
  url          = {https://github.com/anthropics/claude-code/issues/54714}
}

@misc{anthropic2026UsageLimitBestPractices,
  author       = {{Anthropic}},
  title        = {{Usage Limit Best Practices}},
  howpublished = {{Claude Help Center}},
  year         = {2026},
  IGNOREnote         = {Updated over 2 weeks before access; accessed June 19, 2026},
  url          = {https://support.claude.com/en/articles/9797557-usage-limit-best-practices}
}

@article{flanagan_critical_1954,
title   = {The Critical Incident Technique},
author  = {Flanagan, John C.},
journal = {Psychological Bulletin},
IGNOREvolume  = {51},
IGNOREnumber  = {4},
IGNOREpages   = {327–358},
year    = {1954},
IGNOREdoi     = {10.1037/h0061470}
}

@article{butterfield_fifty_2005,
title   = {Fifty Years of the Critical Incident Technique: 1954–2004 and Beyond},
author  = {Butterfield, Lee D. and Borgen, William A. and Amundson, Norman E. and Maglio, Asa-Sophia T.},
journal = {Qualitative Research},
IGNOREvolume  = {5},
IGNOREnumber  = {4},
IGNOREpages   = {475–497},
year    = {2005},
IGNOREdoi     = {10.1177/1468794105056924}
}

@article{gremler_critical_2004,
title   = {The Critical Incident Technique in Service Research},
author  = {Gremler, Dwayne D.},
journal = {Journal of Service Research},
IGNOREvolume  = {7},
IGNOREnumber  = {1},
IGNOREpages   = {65–89},
year    = {2004},
IGNOREdoi     = {10.1177/1094670504266138}
}

@misc{shinn2023reflexion,
  title        = {Reflexion: Language Agents with Verbal Reinforcement Learning},
  author       = {Shinn, Noah and Cassano, Federico and Berman, Edward and Gopinath, Ashwin and Narasimhan, Karthik and Yao, Shunyu},
  year         = {2023},
  eprint       = {2303.11366},
  archivePrefix= {arXiv},
  primaryClass = {cs.AI},
IGNOREdoi          = {10.48550/arXiv.2303.11366},
  url          = {https://arxiv.org/abs/2303.11366}
}

@misc{barke2026agentrx,
  title        = {AgentRx: Diagnosing AI Agent Failures from Execution Trajectories},
  author       = {Barke, Shraddha and Goyal, Arnav and Khare, Alind and Singh, Avaljot and Nath, Suman and Bansal, Chetan},
  year         = {2026},
  eprint       = {2602.02475},
  archivePrefix= {arXiv},
  primaryClass = {cs.AI},
 IGNOREdoi          = {10.48550/arXiv.2602.02475},
  url          = {https://arxiv.org/abs/2602.02475}
}

@misc{ni2026trace2skill,
  title        = {Trace2Skill: Distill Trajectory-Local Lessons into Transferable Agent Skills},
  author       = {Ni, Jingwei and Liu, Yihao and Liu, Xinpeng and Sun, Yutao and Zhou, Mengyu and Cheng, Pengyu and Wang, Dexin and Zhao, Erchao and Jiang, Xiaoxi and Jiang, Guanjun},
  year         = {2026},
  eprint       = {2603.25158},
  archivePrefix= {arXiv},
  primaryClass = {cs.AI},
IGNOREdoi          = {10.48550/arXiv.2603.25158},
  url          = {https://arxiv.org/abs/2603.25158}
}

@misc{lin2026agenticHarnessEngineering,
  title        = {Agentic Harness Engineering: Observability-Driven Automatic Evolution of Coding-Agent Harnesses},
  author       = {Lin, Jiahang and Liu, Shichun and Pan, Chengjun and Lin, Lizhi and Dou, Shihan and Xi, Zhiheng and Huang, Xuanjing and Yan, Hang and Han, Zhenhua and Gui, Tao and Jiang, Yu-Gang},
  year         = {2026},
  eprint       = {2604.25850},
  archivePrefix= {arXiv},
  primaryClass = {cs.CL},
IGNOREdoi          = {10.48550/arXiv.2604.25850},
  url          = {https://arxiv.org/abs/2604.25850}
}

@misc{chen2026harnessfix,
  title        = {From Failed Trajectories to Reliable LLM Agents: Diagnosing and Repairing Harness Flaws},
  author       = {Chen, Mengzhuo and Wang, Junjie and Liu, Zhe and Wang, Yawen and Wang, Qing},
  year         = {2026},
  eprint       = {2606.06324},
  archivePrefix= {arXiv},
  primaryClass = {cs.AI},
 IGNOREdoi          = {10.48550/arXiv.2606.06324},
  url          = {https://arxiv.org/abs/2606.06324}
}

@article{finlay2002reflexivity,

  title     = {Negotiating the Swamp: The Opportunity and Challenge of Reflexivity in Research Practice},
  author    = {Finlay, Linda},
  journal   = {Qualitative Research},
  volume    = {2},
  number    = {2},
  IGNOREpages     = {209--230},
  year      = {2002},
  publisher = {SAGE Publications},
IGNOREdoi       = {10.1177/146879410200200205}

}

@article{tracy2010qualitativeQuality,

  title     = {Qualitative Quality: Eight {``Big-Tent''} Criteria for Excellent Qualitative Research},
  author    = {Tracy, Sarah J.},
  journal   = {Qualitative Inquiry},
  volume    = {16},
  number    = {10},
  year      = {2010},
  publisher = {SAGE Publications},
IGNOREdoi       = {10.1177/1077800410383121}

}

@article{EhsaniDebt2026,
   title={Faster Code, Deeper Debt? A Multivocal Literature Review on Technical Debt and Its Early Signs in LLM-Assisted Software Development},
   ISSN={1557-7392},
   IGNOREurl={http://dx.doi.org/10.1145/3820165},
   IGNOREDOI={10.1145/3820165},
   journal={ACM Transactions on Software Engineering and Methodology},
   publisher={Association for Computing Machinery (ACM)},
   author={Ehsani, Ramtin and Rawal, Shriya and Cai, Yuanfang and Chatterjee, Preetha},
   year={2026},
   month=June
}

@book{yin2018case,
  title     = {Case Study Research and Applications: Design and Methods},
  author    = {Yin, Robert K.},
  edition   = {6},
  year      = {2018},
  publisher = {SAGE Publications},
  address   = {Thousand Oaks, CA}
}

@article{eisenhardt1989building,
  title   = {Building Theories from Case Study Research},
  author  = {Eisenhardt, Kathleen M.},
  journal = {Academy of Management Review},
  year    = {1989}
}

@article{eisenhardt2007theory,
  title   = {Theory Building from Cases: Opportunities and Challenges},
  author  = {Eisenhardt, Kathleen M. and Graebner, Melissa E.},
  journal = {Academy of Management Journal},
  year    = {2007},
}

@book{george2005caseStudies,
  title     = {Case Studies and Theory Development in the Social Sciences},
  author    = {George, Alexander L. and Bennett, Andrew},
  year      = {2005},
  publisher = {MIT Press},
  address   = {Cambridge, MA}
}

@article{seaman1999qualitative,
author={Seaman, C.B.},
  journal={IEEE Transactions on Software Engineering}, 
  title={Qualitative methods in empirical studies of software engineering}, 
  year={1999},
  volume={25},
  number={4},
  pages={557-572},
  doi={10.1109/32.799955}
}

@inproceedings{dittrich2007cooperative,
  title     = {Cooperative Method Development},
  author    = {Dittrich, Yvonne and R{\"o}nkk{\"o}, Kari and Eriksson, Jeanette and Hansson, Christina and Lindeberg, Olof},
  booktitle = {Proceedings of the 2007 ACM GROUP},
  year      = {2007},
IGNOREdoi       = {10.1145/1316624.1316655}
}

@ARTICLE{schafer2024LLMUnitTest,
 author={Schäfer, Max and Nadi, Sarah and Eghbali, Aryaz and Tip, Frank},
  journal={IEEE Transactions on Software Engineering}, 
  title={An Empirical Evaluation of Using Large Language Models for Automated Unit Test Generation}, 
  year={2024},
  volume={50},
  number={1},
  pages={85-105},
  doi={10.1109/TSE.2023.3334955}
}

@inproceedings{sweagent,
author = {Yang, John and Jimenez, Carlos E. and Wettig, Alexander and Lieret, Kilian and Yao, Shunyu and Narasimhan, Karthik and Press, Ofir},
title = {SWE-agent: agent-computer interfaces enable automated software engineering},
year = {2024},
isbn = {9798331314385},
IGNOREpublisher = {Curran Associates Inc.},
IGNOREaddress = {Red Hook, NY, USA},
booktitle = {Proceedings of the 38th International Conference on Neural Information Processing Systems},
articleno = {1601},
numpages = {125},
location = {Vancouver, BC, Canada},
IGNOREseries = {NIPS '24}
}

@inproceedings{zhang-etal-2024-codeagent,
    title = "{C}ode{A}gent: Enhancing Code Generation with Tool-Integrated Agent Systems for Real-World Repo-level Coding Challenges",
    author = "Zhang, Kechi  and
      Li, Jia  and
      Li, Ge  and
      Shi, Xianjie  and
      Jin, Zhi",
    IGNOREeditor = "Ku, Lun-Wei  and
      Martins, Andre  and
      Srikumar, Vivek",
    booktitle = "Proceedings of the 62nd Annual Meeting of the ACL (Volume 1: Long Papers)",
    month = aug,
    year = "2024",
    IGNOREaddress = "Bangkok, Thailand",
    IGNOREpublisher = "Association for Computational Linguistics",
    IGNOREurl = "https://aclanthology.org/2024.acl-long.737/",
    IGNOREdoi = "10.18653/v1/2024.acl-long.737",
    pages = "13643--13658"
}

@inproceedings{autocoderover,
author = {Zhang, Yuntong and Ruan, Haifeng and Fan, Zhiyu and Roychoudhury, Abhik},
title = {AutoCodeRover: Autonomous Program Improvement},
year = {2024},
isbn = {9798400706127},
IGNOREpublisher = {Association for Computing Machinery},
IGNOREaddress = {New York, NY, USA},
IGNOREurl = {https://doi.org/10.1145/3650212.3680384},
IGNOREdoi = {10.1145/3650212.3680384},
booktitle = {Proceedings of the 33rd ACM SIGSOFT ISSTA},
IGNOREpages = {1592–1604},
numpages = {13},
location = {Vienna, Austria},
IGNOREseries = {ISSTA 2024}
}

@misc{deng2025swebenchproaiagents,
      title={SWE-Bench Pro: Can AI Agents Solve Long-Horizon Software Engineering Tasks?}, 
      author={Xiang Deng and Jeff Da and Edwin Pan and Yannis Yiming He and Charles Ide and Kanak Garg and Niklas Lauffer and Andrew Park and Nitin Pasari and Chetan Rane and Karmini Sampath and Maya Krishnan and Srivatsa Kundurthy and Sean Hendryx and Zifan Wang and Vijay Bharadwaj and Jeff Holm and Raja Aluri and Chen Bo Calvin Zhang and Noah Jacobson and Bing Liu and Brad Kenstler},
      year={2025},
      eprint={2509.16941},
      archivePrefix={arXiv},
      primaryClass={cs.SE},
      url={https://arxiv.org/abs/2509.16941}, 
}

@misc{ding2026nl2repobenchlonghorizonrepositorygeneration,
      title={NL2Repo-Bench: Towards Long-Horizon Repository Generation Evaluation of Coding Agents}, 
      author={Jingzhe Ding and Shengda Long and Changxin Pu and Huan Zhou and Hongwan Gao and Xiang Gao and Chao He and Yue Hou and Fei Hu and Zhaojian Li and Weiran Shi and Zaiyuan Wang and Daoguang Zan and Chenchen Zhang and Xiaoxu Zhang and Qizhi Chen and Xianfu Cheng and Bo Deng and Qingshui Gu and Kai Hua and Juntao Lin and Pai Liu and Mingchen Li and Xuanguang Pan and Zifan Peng and Yujia Qin and Yong Shan and Zhewen Tan and Weihao Xie and Zihan Wang and Yishuo Yuan and Jiayu Zhang and Enduo Zhao and Yunfei Zhao and He Zhu and Liya Zhu and Chenyang Zou and Ming Ding and Jianpeng Jiao and Jiaheng Liu and Minghao Liu and Qian Liu and Chongyang Tao and Jian Yang and Tong Yang and Zhaoxiang Zhang and Xinjie Chen and Wenhao Huang and Ge Zhang},
      year={2026},
      eprint={2512.12730},
      archivePrefix={arXiv},
      primaryClass={cs.CL},
      url={https://arxiv.org/abs/2512.12730}, 
}

@misc{yang2026programbenchlanguagemodelsrebuild,
      title={ProgramBench: Can Language Models Rebuild Programs From Scratch?}, 
      author={John Yang and Kilian Lieret and Jeffrey Ma and Parth Thakkar and Dmitrii Pedchenko and Sten Sootla and Emily McMilin and Pengcheng Yin and Rui Hou and Gabriel Synnaeve and Diyi Yang and Ofir Press},
      year={2026},
      eprint={2605.03546},
      archivePrefix={arXiv},
      primaryClass={cs.SE},
      url={https://arxiv.org/abs/2605.03546}, 
}

@inproceedings{sinha2025illusion,
    title={The Illusion of Diminishing Returns: Measuring Long Horizon Execution in {LLM}s},
    author={Akshit Sinha and Arvindh Arun and Shashwat Goel and Steffen Staab and Jonas Geiping},
    booktitle={The Fourteenth International Conference on Learning Representations},
    year={2026},
}

@article{bjornson2008knowledge,
  title={Knowledge management in software engineering: A systematic review of studied concepts, findings and research methods used},
  author={Bj{\o}rnson, Finn Olav and Dings{\o}yr, Torgeir},
  journal={Information and software technology},
  year={2008},
  publisher={Elsevier}
}

@article{gao2025SELLMChallenges,
author = {Gao, Cuiyun and Hu, Xing and Gao, Shan and Xia, Xin and Jin, Zhi},
title = {The Current Challenges of Software Engineering in the Era of Large Language Models},
year = {2025},
issue_date = {June 2025},
publisher = {Association for Computing Machinery},
address = {New York, NY, USA},
volume = {34},
number = {5},
issn = {1049-331X},
url = {https://doi.org/10.1145/3712005},
doi = {10.1145/3712005},
journal = {ACM Trans. Softw. Eng. Methodol.},
month = may,
articleno = {127},
numpages = {30}
}

@inproceedings{sarkar2025vibecoding,
author={Sarkar, Advait and Drosos, Ian}, booktitle={{Proceedings of the 36th Annual Conference of the Psychology of Programming Interest Group
(PPIG 2025)}}, title={{Vibe coding: programming through conversation with artificial intelligence}}, month=sep, year={2025}}

@inproceedings{kirova2024SEShouldAdaptLLM,
author = {Kirova, Vassilka D. and Ku, Cyril S. and Laracy, Joseph R. and Marlowe, Thomas J.},
title = {Software Engineering Education Must Adapt and Evolve for an LLM Environment},
year = {2024},
isbn = {9798400704239},
IGNOREpublisher = {Association for Computing Machinery},
IGNOREaddress = {New York, NY, USA},
url = {https://doi.org/10.1145/3626252.3630927},
doi = {10.1145/3626252.3630927},
booktitle = {Proceedings of the 55th ACM Technical Symposium on Computer Science Education V. 1},
pages = {666–672},
numpages = {7},
location = {Portland, OR, USA},
IGNOREseries = {SIGCSE 2024}
}

@ARTICLE{randall2024AISECurriculum,
  author={Randall, Natasha and Wäckerle, Dennis and Stein, Nils and Goßler, Dennis and Bente, Stefan},
  journal={IEEE Software}, 
  title={What an AI-Embracing Software Engineering Curriculum Should Look Like: An Empirical Study}, 
  year={2024},
  volume={41},
  number={2},
  pages={36-43},
  doi={10.1109/MS.2023.3344682}}

@article{agarwal2026ai,
      title={AI IDEs or Autonomous Agents? Measuring the Impact of Coding Agents on Software Development}, 
      author={Shyam Agarwal and Hao He and Bogdan Vasilescu},
      year={2026},
      eprint={2601.13597},
      archivePrefix={arXiv},
      primaryClass={cs.SE},
      url={https://arxiv.org/abs/2601.13597}, 
}

@misc{thawar2026whatjobnow,
  author       = {Thawar, Farhan},
  title        = {{What Is Your Job Now, Farhan Thawar \textbar{} Compile 26}},
  year         = {2026},
  howpublished = {YouTube video},
  IGNOREnote         = {Cursor account. Accessed: 2026-07-01},
  url          = {https://www.youtube.com/watch?v=ByOF8qByGHU}
}

@incollection{popper2012science,
  title={Science: Conjectures and refutations},
  author={Popper, Karl},
  booktitle={Arguing about Science},
  pages={15--43},
  year={2012},
  publisher={Routledge}
}

@inproceedings{peng2026beyond,
  title={Beyond Local Code Optimization: Multi-Agent Reasoning for Software System Optimization},
  author={Peng, Huiyun and Patil, Parth Vinod and Qiu, Antonio Zhong and Thiruvathukal, George K and Davis, James C},
  booktitle={[ICSE-JAWs'26] The First Journal Ahead Workshop (co-located with ICSE'26)},
  year={2026}
}

@inproceedings{kwa2026measuring,
    title={Measuring {AI} Ability to Complete Long Software Tasks},
    author={Thomas Kwa and Ben West and Joel Becker and Amy Deng and Katharyn Garcia and Max Hasin and Sami Jawhar and Megan Kinniment and Nate Rush and Sydney Von Arx and Ryan Bloom and Thomas Broadley and Haoxing Du and Brian Goodrich and Nikola Jurkovic and Luke Harold Miles and Seraphina Nix and Tao Roa Lin and Neev Parikh and David Rein and Lucas Jun Koba Sato and Hjalmar Wijk and Daniel M Ziegler and Elizabeth Barnes and Lawrence Chan},
    booktitle={The Thirty-ninth Annual Conference on Neural Information Processing Systems},
    year={2026},
}

@inproceedings{anandayuvaraj2026learning,
  title={Learning From Software Failures: A Case Study at a National Space Research Center},
  author={Anandayuvaraj, Dharun and Hammadeh, Zain and Lund, Andreas and Holloway, Alexandra and Davis, James C},
  booktitle={[ICSE'26] Proceedings of the 48th IEEE/ACM International Conference on Software Engineering},
  year={2026}
}

\end{document}